\documentclass{aa}
\usepackage{graphics}

\begin{document}      

   \title{Pre-peak ram pressure stripping in the Virgo cluster spiral galaxy NGC~4501}

   \author{B.~Vollmer\inst{1}, M.~Soida\inst{2}, A.~Chung\inst{3}, J.H.~van~Gorkom\inst{4}, 
     K.~Otmianowska-Mazur\inst{2}, R.~Beck\inst{5}, M.~Urbanik\inst{2}, J.D.P.~Kenney\inst{6}}

   \offprints{B.~Vollmer, e-mail: bvollmer@astro.u-strasbg.fr}

   \institute{CDS, Observatoire astronomique de Strasbourg, 11, rue de l'universit\'e,
	      67000 Strasbourg, France \and
	      Astronomical Observatory, Jagiellonian University, Orla 171,
	      30-244 Krak\'ow, Poland \and
	      NRAO Jansky fellow at University of Massachusetts, Amherst, MA 01003, USA, \and
	      Department of Astronomy, Columbia University, 538 West 120th Street, New York, 
	      NY 10027, USA \and
	      Max-Planck-Insitut f\"{u}r Radioastronomie, Auf dem H\"{u}gel 69, 53121 Bonn, Germany \and
	      Yale University Astronomy Department, P.O. Box 208101, New Haven, CT 06520-8101, USA
              }

   \date{Received / Accepted}

   \authorrunning{Vollmer et al.}
   \titlerunning{Pre-peak ram pressure stripping in NGC~4501}

\abstract{
VIVA H{\sc i} observations of the Virgo spiral galaxy NGC~4501 are presented.
The H{\sc i} disk is sharply truncated to the southwest, well within the stellar disk.
A region of low surface-density gas, which is more extended than the
main H{\sc i} disk, is discovered northeast of the galaxy center.
These data are compared to existing 6~cm polarized radio continuum emission,
H$\alpha$, and optical broad band images. We observe a coincidence between the
western H{\sc i} and polarized emission edges, on the one hand, and a faint
H$\alpha$ emission ridge, on the other. The polarized emission maxima are located
within the gaps between the spiral arms and the faint H$\alpha$ ridge. 
Based on the comparison of these observations with
a sample of dynamical simulations with different values for maximum
ram pressure and different inclination angles between the disk and the orbital plane,
we conclude that ram pressure stripping can account for the main observed characteristics.
NGC~4501 is stripped nearly edge-on, is heading southwest, and is 
$\sim 200-300$~Myr before peak ram pressure, i.e. its closest approach to M87.
The southwestern ridge of enhanced gas surface density and enhanced
polarized radio-continuum emission is due to ram pressure compression. 
It is argued that the faint western H$\alpha$ emission ridge
is induced by nearly edge-on ram pressure stripping. NGC~4501 represents an
especially clear example of early stage ram pressure stripping of a 
large cluster-spiral galaxy.
}

\maketitle

\section{Introduction \label{sec:intro}}

The Virgo cluster is dynamically young and spiral-rich
making it an ideal laboratory for studying the influence of the cluster environment
on spiral galaxies. Most of the Virgo spiral galaxies are H{\sc i} deficient,
i.e. they have lost a significant amount of their ISM 
(Chamaraux et al. 1980, Giovanelli \& Haynes 1983). 
Imaging H{\sc i} observations have shown that 
these galaxies have truncated H{\sc i} disks (Cayatte et al. 1990).
Thus, the cluster environment changes the H{\sc i} content and morphology of
Virgo cluster spiral galaxies. However, it remains unclear at which
cluster distance the spiral galaxies lose their external H{\sc i} disk.
Based on deep Effelsberg H{\sc i} observations around 6 Virgo galaxies
and a balance of previous detections of extraplanar gas in targeted galaxies 
Vollmer \& Huchtmeier (2007) propose a global picture where the outer gas disk 
(beyond the optical radius $R_{25}$) is removed much earlier than expected by 
the classical ram pressure criterion. They argue that the vulnerable outer disk
is removed much more rapidly than predicted by the Gunn \& Gott criterion.
Furthermore, Chung et al. (2007) found in a new H{\sc i} imaging survey of
$\sim 50$  Virgo spiral galaxies 7 galaxies with long H{\sc i} tails pointing away from M87.
These galaxies are located at intermediate distances from the cluster center (0.6--1~Mpc).
They conclude that these one-sided H{\sc i} tail galaxies have recently arrived in the cluster, 
falling in on highly radial orbits. It appears that galaxies begin to lose their gas 
already at intermediate distances from the cluster center through ram-pressure or 
turbulent viscous stripping (Nulsen 1982) and tidal interactions with neighbours, 
or a combination of both. The outer gas disks of spiral galaxies at projected distances 
smaller than $\sim1$~Mpc are thus affected by ram pressure. In this article we show that this
is also the case for the massive spiral galaxy NGC~4501 (for its properties see 
table~\ref{tab:parameters}).

NGC~4501 is located at the lower end of the distance interval defined by Chung et al. (2007).
Moreover, Cayatte et al. (1990, 1994) and B\"{o}hringer et al. (1997) proposed that
it approaches the center of the Virgo cluster (M87) on a highly eccentric orbit.
NGC~4501 is moderately H{\sc i} deficient and has an H{\sc i} disk whose southwestern side
is truncated well within optical radius ($R_{25}$). The H{\sc i} surface density distribution 
within the disk is asymmetric with an overdense region to the southwest
showing a sharp outer edge (Cayatte et al. 1990). 
Since the direction of these features is pointing toward the Virgo
cluster center, Cayatte et al. (1990, 1994) and B\"{o}hringer et al. (1997) 
speculate that they are due to ram pressure compression (Gunn \& Gott 1972).
Since ram pressure only affects the interstellar medium of the galaxy and
not its stars, one expects a symmetric stellar disk. However, the 
DSS B band morphology is asymmetric, the southwestern side of the major
axis being brighter than the northeastern side, rising the question if NGC~4501
is exclusively experiencing an interaction with the intracluster medium
and if a past gravitational interaction can be excluded.

In this article we investigate the nature of the interaction of NGC~4501 
with the cluster environment using VIVA (VLA Imaging of Virgo galaxies in Atomic gas,
Chung et al. in prep.) H{\sc i} data, VLA 6~cm polarized radio emission data, and a
detailed dynamical model including MHD calculations.
Whereas the H{\sc i} distribution and kinematics show a distorted gas
disk where some of the gas is pushed to the north-east, the polarized
radio-continuum emission represents an ideal tracer of ram pressure compression
at the opposite side of the disk.
The comparison between observations and simulations shows that
NGC~4501 is indeed approaching the cluster center. Ram pressure
stripping can account for the main properties of this Virgo spiral galaxy.

We describe the H{\sc i} and 6~cm polarized radio-continuum observations and their results 
in Sect.~\ref{sec:observations} and \ref{sec:results}. Additional H$\alpha$ data are discussed
in Sect.~\ref{sec:starform}. After a description of the dynamical model, the best fitting model 
and following MHD calculations are presented (Sect.~\ref{sec:model}). Our models
are compared to the observational data in Sect.~\ref{sec:comparison} followed by a
discussion (Sect.~\ref{sec:discussion}) and our conclusions (Sect.~\ref{sec:conclusions}).

\begin{table}
      \caption{Physical Parameters of NGC~4501}
         \label{tab:parameters}
      \[
         \begin{array}{lr}
            \hline
            \noalign{\smallskip}
        {\rm Other\ names} &  {\rm M~88} \\
                & {\rm VCC~1401} \\
                & {\rm UGC~7675}  \\
        $$\alpha$$\ (2000)$$^{\rm a}$$ &  12$$^{\rm h}31^{\rm m}59.2^{\rm s}$$\\
        $$\delta$$\ (2000)$$^{\rm a}$$ &  14$$^{\rm o}25'14''$$\\
        {\rm Morphological\ type}$$^{\rm a}$$ & {\rm Sb} \\
        {\rm Distance\ to\ the\ cluster\ center} & 2.0$$^{\rm o},\ 0.6~{\rm Mpc}$$\\
        {\rm Optical\ diameter\ D}_{25}$$^{\rm a}$$ & 6.9$$'$$,\ 34~{\rm kpc} \\
        {\rm B}$$_{T}^{0}$$$$^{\rm a}$$ & 9.86\\ 
        {\rm Systemic\ heliocentric\ velocity}$$^{\rm a}$$\ {\rm (km\,s}$$^{-1}$$)\ & 2280$$\pm$$4\\
	{\rm Velocity\ with\ respect\ to\ Virgo\ mean}\  {\rm (km\,s}$$^{-1}$$)\ & 1130\\
        {\rm Distance\ D\ (Mpc)} & 17 \\
        {\rm Vrot}$$_{\rm max}\ {\rm (km\,s}$$^{-1}$$) & 300$$^{\rm b}$$ \\
        {\rm PA} & 142$$^{\rm o}$$\ $$^{\rm b}$$\\
        {\rm Inclination\ angle} & 56$$^{\rm o}$$\ $$^{\rm b}$$ \\
        {\rm HI\ deficiency}^{\rm c}$$ &  0.56$$\pm$$0.2\\
        \noalign{\smallskip}
        \hline
        \end{array}
      \]
\begin{list}{}{}
\item[$^{\rm{a}}$] RC3, de Vaucouleurs et al. (1991)
\item[$^{\rm{b}}$] Guharthakurta et al. (1988)
\item[$^{\rm{c}}$] Cayatte et al. (1994)
\end{list}
\end{table}

\section{Observations \label{sec:observations}}

\subsection{VIVA H{\sc i} 21~cm line observations}

H{\sc i} line observations were made on January 20 1999, using the C configuration 
of the VLA of the National Radio Astronomy Observatory (NRAO)\footnote{NRAO is a facility 
of National Science Foundation
operated under cooperative agreement by Associated Universities, Inc.}. 
The total on-source time was
5.6 hr. The bandpasses of 2$\times$3.125~MHz were centered at 2280~km~s$^{-1}$ 
configured to produce two polarizations and 63 channels
with on-line Hanning smoothing, yielding a channel spacing of 10.4~km~s$^{-1}$.
The nearby phase calibrator 1252+199
was observed every 25 minutes, and 1328+307 was used as a flux calibrator.

We used the NRAO Astronomical Image Processing System (AIPS) for calibration and
mapping. The continuum subtraction was done using a linear fit to the visibility
data of line-free channels. To make the image cube we applied a weighting scheme
intermediate between uniform and natural, but closer to a natural weighting scheme
(robust=1), which resulted in a resolution of $17''$. To maximize sensitivity, 
we made a second data cube with tapered data (resolution of $30''$). 
The resulting rms are 0.7~mJy and 1.2~mJy per beam and per channel, respectively,
which corresponds to $2.9/1.5 \times 10^{19}$~cm$^{-2}$.

\subsection{6~cm polarized radio-continuum observations}

NGC~4501 was observed for 3:55~h on December 3 2005 with the Very Large Array 
(VLA) of the National Radio 
Astronomy Observatory (NRAO) in the D array configuration. The band passes were $2\times 50$~MHz.
We used 3C286 as the flux calibrator and 1254+116 as the phase calibrator, the latter of
which was observed every 40~min. 
Maps were made for both wavelengths using the AIPS task IMAGR with ROBUST=3.
The final cleaned maps were convolved to a beam size of $18'' \times 18''$.
We ended up with an rms level of the linear polarization, taken to be the mean rms
in Stokes Q and U, of $11$~$\mu$Jy/beam. The distribution of 6~cm polarized continuum 
emission was already published in Vollmer et al. (2007).
For each map point we determined the apparent polarization B-vector, defined as the
E-vector rotated by $90^{\circ}$. Taking into account the possible bias due to
Faraday rotation (We\.zgowiec et al. 2007) this represents a good approximation
of the orientation of the sky-projected mean regular magnetic field, accurate to
about $\pm 10^{\circ}$.

\section{Results \label{sec:results}}

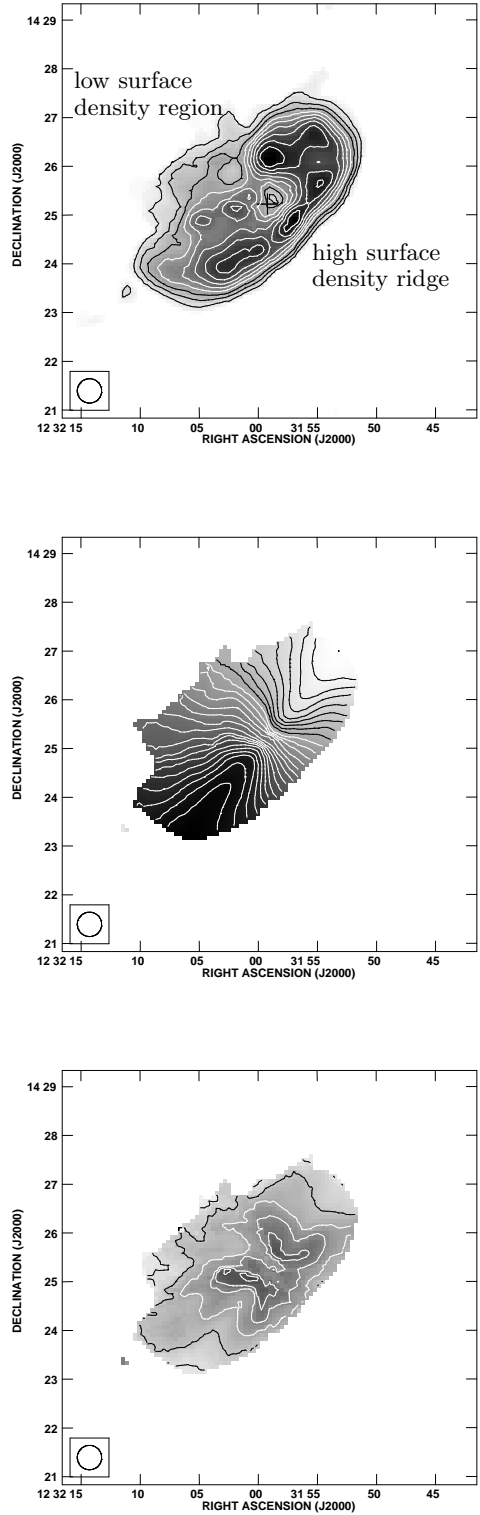
\begin{figure} 
	\resizebox{7cm}{!}{\includegraphics{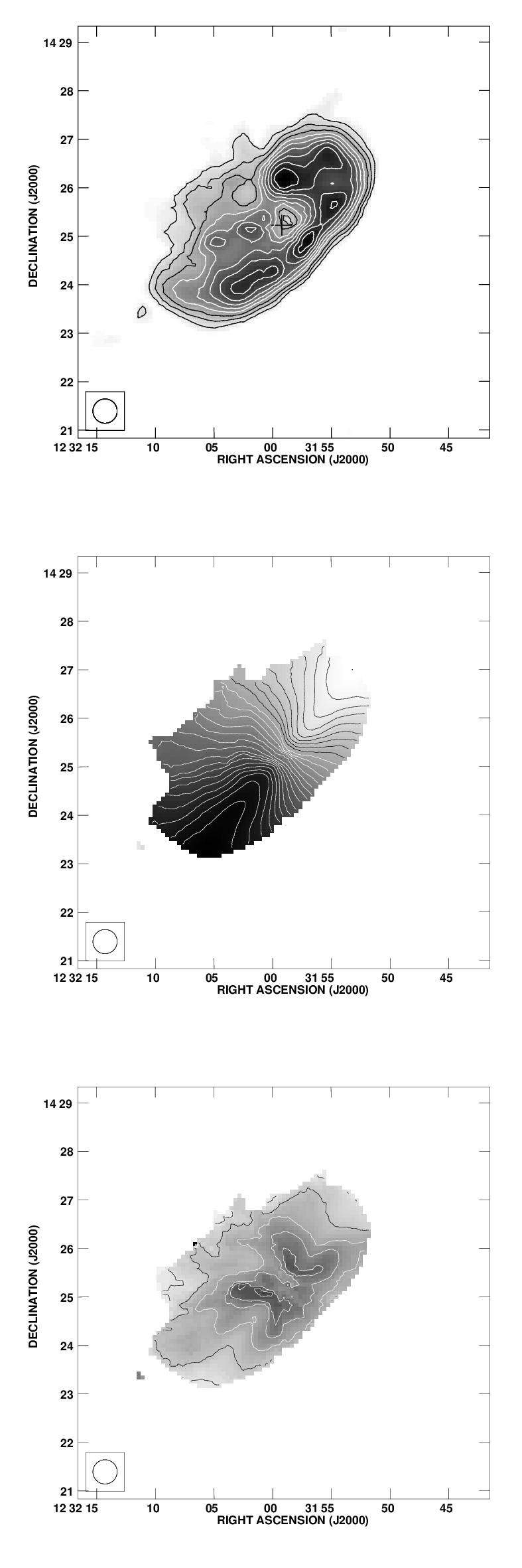}}
	\put(-165,555){low surface}
	\put(-165,545){density region}
	\put(-75,490){high surface}
	\put(-75,480){density ridge}
	\caption{H{\sc i} observations of NGC~4501.
	  Upper panel: surface density distribution. Contour levels
	  are (1, 2, 3, 4, 5, 6, 7, 8, 9, 10)$\times$(0.125~Jy/beam\,km\,s$^{-1}$
	  or $1.5 \times 10^{20}$~cm$^{-2}$).
	  The beam ($30'' \times 30''$) is plotted 
	  in the lower left corner of the image.
	  Middle panel: velocity field. Contours are from 2010~km\,s$^{-1}$ (northwest)
	  to 2590~km\,s$^{-1}$ (southeast) in steps of 20~km\,s$^{-1}$.
	  Lower panel: velocity dispersion distribution.
	  Contour levels are (10, 20, 30, 40, 50)~km\,s$^{-1}$.
	} \label{fig:HImoments}
\end{figure}

\begin{figure} 
	\resizebox{\hsize}{!}{\includegraphics{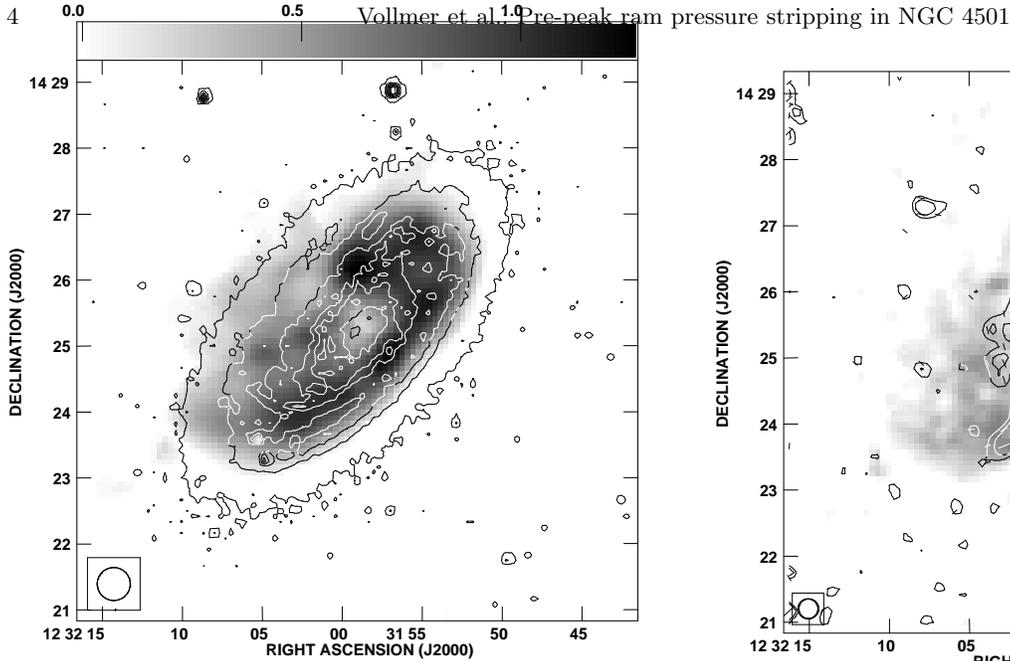}}
	\caption{Greyscale: H{\sc i} surface density distribution 
	  (0 to 1.25 Jy/beam\,km\,s$^{-1}$; resolution: $30''$). 
	  Contours: B band image from GOLD Mine (Gavazzi et al. 2003).
	} \label{fig:HIB}
\end{figure}

\begin{figure} 
	\resizebox{\hsize}{!}{\includegraphics{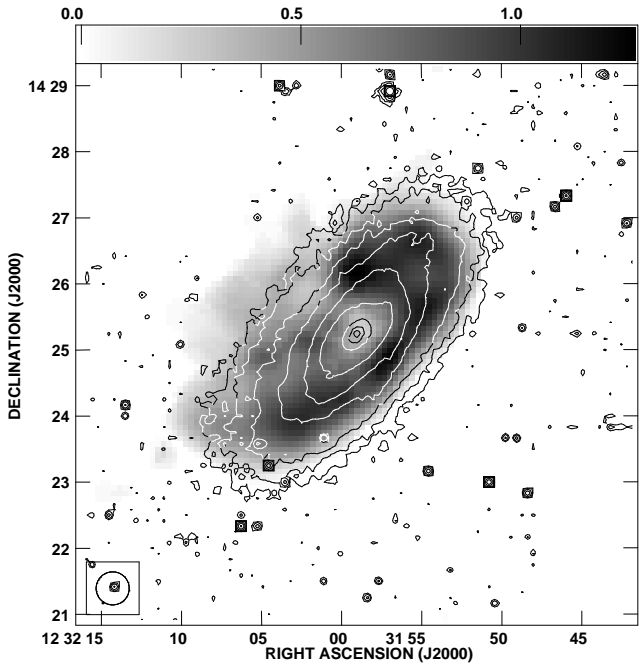}}
	\caption{Greyscale: H{\sc i} surface density distribution 
	  (0 to 1.25 Jy/beam\,km\,s$^{-1}$; resolution: $30''$).
	  Contours: H band image from GOLD Mine (Gavazzi et al. 2003).
	} \label{fig:HIH}
\end{figure}

The $30''$ resolution H{\sc i} distribution, velocity field, and velocity dispersion distribution are
shown in Fig.~\ref{fig:HImoments}. The superposition of the H{\sc i} distribution on
a B band and an H band (1.65~$\mu$m) image are presented in Figs.~\ref{fig:HIB} and \ref{fig:HIH}.
As observed by Cayatte et al. (1990) the H{\sc i} distribution shows a central hole and
is truncated near the optical radius $R_{25}$. 
The H{\sc i} surface density distribution within the disk
is asymmetric with a high surface-density ridge in the southwest. The average H{\sc i}
surface density in the northeastern part of the disk is $\sim$3 times lower
than that of the southwestern ridge. We observe low surface-density H{\sc i}
($\sim 2-5 \times 10^{19}$~cm$^{-2}$) to the northeast of the galaxy center. 
This gas was not detected by Cayatte et al. (1990).

The velocity field of NGC~4501 is relatively symmetric parallel to the minor axis
even in the low surface-density
northeastern region. The rotation curve becomes flat at a smaller galactocentric
radius in the southeast. The velocity dispersion is symmetric with two maxima
caused by beam smearing over the inner part of the galaxy with a steeply rising
rotation curve.

The comparison between the H{\sc i} surface density distribution and
optical images (Fig.~\ref{fig:HIB} and \ref{fig:HIH}) shows that
the southwestern part of the H{\sc i} disk is sharply truncated
to well within the stellar disk.
The outermost H{\sc i} disk follows one isophotal contour in the southwest,
but extends well beyond this contour in the northeast.
This is also the case for a small ($\sim 20''$)
straight low surface-density region in the north pointing to the northwest. 
As already mentioned in Sect.~\ref{sec:intro}, NGC~4501 has an asymmetric
surface brightness distribution in the B band with flocculent spiral arms 
showing pitch angles which are larger in the northeastern part of the galactic disk
than in the southwestern part. In addition, the southwestern part of the 
disk is brighter than the northeastern part. However, the
last B band isophot contour is symmetric. In addition, NGC~4501
has a symmetric surface brightness distribution in the H band which
shows the old stellar component. We thus conclude that the B band
asymmetries are due to the star formation history and do not reflect
an asymmetry of the gravitational potential. 

\begin{figure} 
	\resizebox{\hsize}{!}{\includegraphics{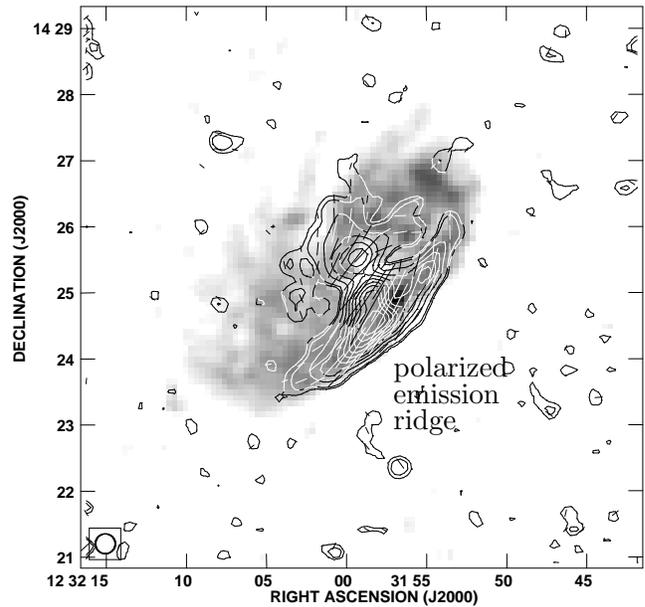}}
	\put(-100,90){\large polarized}
	\put(-100,80){\large emission}
	\put(-100,70){\large ridge}
	\caption{Greyscale: H{\sc i} surface density distribution 
	  (0 to 0.65 Jy/beam\,km\,s$^{-1}$; resolution: $17''$). Contours:
	  6~cm polarized radio-continuum emission (Vollmer et al. 2007).
	  The contour levels are (0.7, 1, 2, 3, 4, 5, 6, 7, 8, 9, 10)$\times$
	  72~$\mu$Jy/beam. The vectors of the magnetic field are uncorrected for
	  Faraday rotation. Their size is proportional to the intensity 
	  of the polarized emission. The beam ($18'' \times 18''$) is plotted 
	  in the lower left corner of the image.	  
	} \label{fig:HIPI}
\end{figure}

\begin{figure} 
	\resizebox{\hsize}{!}{\includegraphics{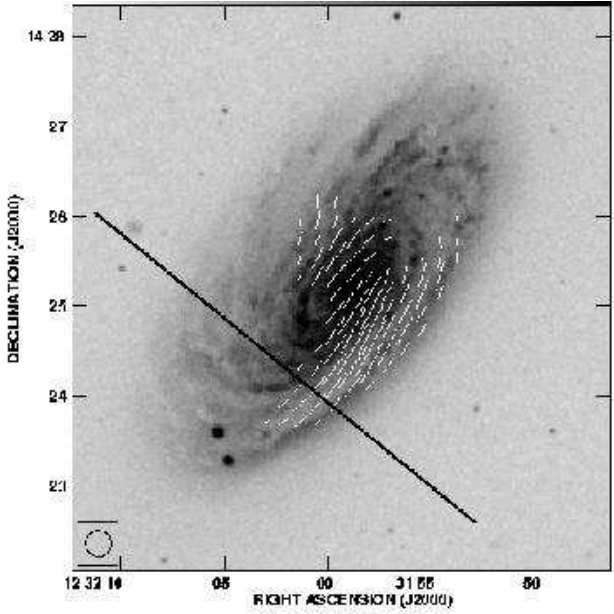}}
	\caption{Greyscale: B band image from GOLD Mine (Gavazzi et al. 2003).
	  The vectors of the magnetic field are uncorrected for
	  Faraday rotation (resolution: $18''$). Their size is proportional to the intensity 
	  of the polarized emission as in Fig.~\ref{fig:HIPI}.
	  The line indicates the slit position for the profile of Fig.~\ref{fig:profiles}.
	} \label{fig:PIB}
\end{figure}

\begin{figure} 
	\resizebox{\hsize}{!}{\includegraphics{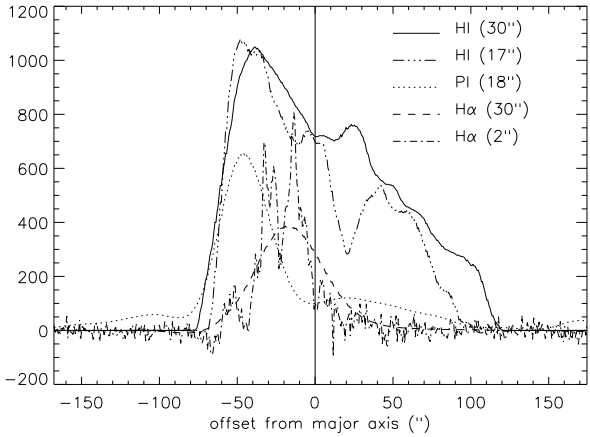}}
	\caption{Profiles parallel to the minor axis $\sim 1'$ southeast of the galaxy center
	  (see Fig.~\ref{fig:PIB}).
	  Solid: H{\sc i} surface density (resolution: $30''$);
	  dash-double-dotted: H{\sc i} surface density (resolution: $17''$);
	  dotted: 6~cm polarized radio-continuum emission (resolution: $18''$);
	  dashed: H$\alpha$ emission at a resolution of $30''$; dash-dotted: H$\alpha$
	  emission at a $2''$ resolution. The $y$-axis has arbitrary units.
	  Left corresponds to the southwest. The position of the major axis is marked
	  with the vertical line.
	} \label{fig:profiles}
\end{figure}

The comparison between the 6~cm polarized radio-continuum emission distribution 
and the $17''$ resolution H{\sc i} surface density distribution is shown in Fig.~\ref{fig:HIPI}.
The polarized emission shows an enhanced ridge south of the galaxy center.
This ridge coincides with the high column density H{\sc i} ridge and
is located entirely within the H{\sc i} gas distribution.
In addition, we observe a local maximum of polarized emission $\sim 20''$ to the north
of the galaxy center.
In normal field spirals the polarized radio-continuum emission is found in 
interarm regions, with B-vectors parallel to the adjacent spiral arms (Beck 
2005). As this inner interarm region is not resolved in our radio polarization 
data, the origin of the polarized emission in the north remains unclear.
The magnetic field vectors are compared to the optical B band images in
Fig.~\ref{fig:PIB}. If the large-scale magnetic field is determined by the spiral arms,
the magnetic field vectors should follow the spiral structure.
This is what we observe in the inner $\sim 1'= 5$~kpc.
However, the magnetic field vectors cross the spiral arm in the most northeastern
part of the polarized emission distribution. This is similar to the
behavior of the large-scale magnetic field in NGC~3627 (Soida et al. 2001) where a
magnetic arm crosses an optical arm, which is interpreted as 
a sign of tidal interactions within the Leo Triplet group. 
On the other hand, the magnetic field within the
ridge of enhanced polarized radio-continuum emission follows the spiral structure.
Fig.~\ref{fig:profiles} shows profiles parallel to the minor axis
centered on the major axis $\sim 1'$ southeast of the galaxy center (see Fig.~\ref{fig:PIB}). 
We show the H{\sc i} surface density,
the 6~cm polarized radio-continuum emission, and the H$\alpha$ emission.
The H{\sc i} and the polarized emission profiles rise together and peak at the same location.
The polarized radio-continuum emission is enhanced by ram pressure
compression of the magnetic field. 
If the original magnetic field was random, the compressed 
field is anisotropic and we do not expect Faraday rotation. 
A detailed analysis of Faraday rotation measures of Virgo Cluster spirals
will be subject of future studies. The intensity 
ratio between the peak and the base of the polarized emission can be used to measure 
the compression ratio (Beck et al. 2005).
The  high resolution faint H$\alpha$ emission coincides with the H{\sc i} and
polarized emission peaks.
A similar behavior is also observed in the interacting spiral galaxy NGC~2276 
(Hummel \& Beck 1995) where a ridge of enhanced polarized radio-continuum emission
is observed in the south of the galaxy, in the group spiral galaxy NGC~2442 (Harnett et al. 2004) 
where the northern polarized emission ridge is shifted towards the outer edge of the star-forming 
region, and the Virgo spiral galaxy NGC~4254 (Soida et al. 1996) where the 
H$\alpha$-bright spiral arm is placed inwards with respect to the compressed region.
In NGC~2276 a tidal compression due 
to a galaxy-galaxy interaction is responsible for the enhanced polarized emission.
For NGC~2442 and NGC~4254 a tidal interaction also cannot be excluded.

\section{Star formation \label{sec:starform}}

\begin{figure} 
	\resizebox{\hsize}{!}{\includegraphics{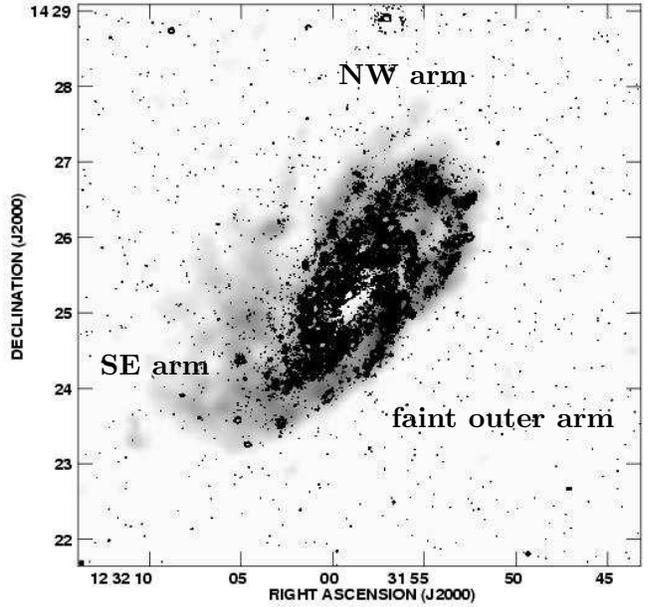}}
	\put(-100,70){\large \bf faint outer arm}
	\put(-120,200){\large \bf NW arm}
	\put(-210,90){\large \bf SE arm}
	\caption{Greyscale: H{\sc i} surface density distribution
	  (0 to 0.65~Jy/beam km\,s$^{-1}$; $17''$ resolution).
	  Contours: H$\alpha$ emission distribution from GOLD Mine (Gavazzi et al. 2003).
	  The contour levels are chosen in a way to make the faint southwestern
	  H$\alpha$ arm visible.
	} \label{fig:HIHA}
\end{figure}

\begin{figure} 
	\resizebox{\hsize}{!}{\includegraphics{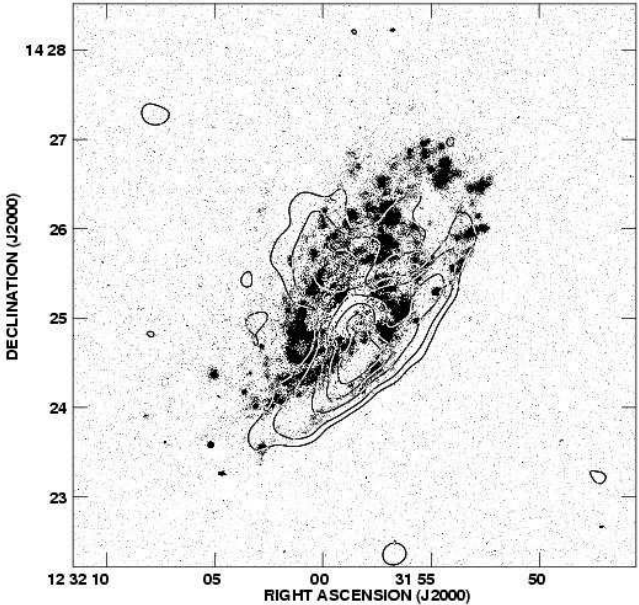}}
	\caption{Greyscale: H$\alpha$ emission distribution. Contours:
	  6~cm polarized radio-continuum emission.
	  The contour levels are (1, 2, 4, 6, 8, 10)$\times$
	  72~$\mu$Jy/beam.
	} \label{fig:PIHA}
\end{figure}

We use an H$\alpha$ image to compare the H{\sc i} and polarized radio-continuum emission
to the distribution of massive star formation (Figs.~\ref{fig:HIHA} and \ref{fig:PIHA}).
The H$\alpha$ distribution shows a ring structure in the inner disk ($R < 40''$) and a prominent
spiral arm in the northwest. The spiral arm begins north of the galaxy center,
continues to the northwest, bends to the southwest, and finally follows the 
southwestern edge of the H{\sc i} distribution. There is the beginning of a
second H$\alpha$ spiral arm in the southeast of the galactic disk. This second
spiral arm abruptly ends $\sim 1'=5$~kpc from the galaxy center.
The comparison between the H$\alpha$ emission distribution and the 6~cm
polarized radio-continuum emission distribution (Fig.~\ref{fig:PIHA})
shows that the maximum of the polarized emission south of the galaxy center
is located between the inner H$\alpha$ ring and the outer H$\alpha$ arm. 
The location of maximum polarized radio-continuum emission may indicate the 
region of strongest field compression, coinciding with maximum HI emission. 
Alternatively, the field could be tangled in star forming regions located in 
the H$\alpha$ arms.
Note that the limited resolution of the radio observations leads to beam 
depolarization in the inner disk around the major axis if the field lines 
follow the strong curvature of the optical spiral structure.
The second local maximum northwest from the first maximum is also located in
an H$\alpha$ emission gap. On the other hand, the two local maxima 
located at $\sim 20''$ north and south of the galaxy center do not coincide 
with such clear gaps. The maxima in the ridge of enhanced polarized emission
are thus different from those of the inner disk, which are possibly due to the shear
caused by the spiral arms. Spitzer $8$~$\mu$m and $24$~$\mu$m maps
(in prep.) show essentially the same spiral arm features as the H$\alpha$.

\section{Models \label{sec:model}}

To interpret the observations described in Sect.~\ref{sec:results}, we  
compare them to a detailed dynamical model which includes ram pressure.
In a second step, we solve the induction equation for the velocity fields
of the dynamical model to determine the evolution of the large-scale
magnetic field. Assuming a distribution of relativistic electrons, we
then calculate the model distribution of polarized radio-continuum emission.

\subsection{Dynamical model}

We use the N-body code described in Vollmer et al. (2001) which consists of 
two components: a non-collisional component
that simulates the stellar bulge/disk and the dark halo, and a
collisional component that simulates the ISM.

The non--collisional component consists of 49\,125 particles, which simulate
the galactic halo, bulge, and disk.
The characteristics of the different galactic components are shown in
Table~\ref{tab:param}.
\begin{table}
      \caption{Total mass, number of particles $N$, particle mass $M$, and smoothing
        length $l$ for the different galactic components.}
         \label{tab:param}
      \[
         \begin{array}{lllll}
           \hline
           \noalign{\smallskip}
           {\rm component} & M_{\rm tot}\ ({\rm M}$$_{\odot}$$)& N & M\ ({\rm M}$$_{\odot}$$) & l\ ({\rm pc}) \\
           \hline
           {\rm halo} & 7.3\,10$$^{11}$$ & 16384 & $$4.5\,10^{7}$$ & 1200 \\
           {\rm bulge} & 2.5\,10$$^{10}$$ & 16384 & $$1.5\,10^{6}$$ & 180 \\
           {\rm disk} & 1.3\,10$$^{11}$$ & 16384 & $$7.9\,10^{5}$$ & 240 \\
           \noalign{\smallskip}
        \hline
        \end{array}
      \]
\end{table}
The resulting rotation velocity is $\sim$300~km\,s$^{-1}$. 

We have adopted a model where the ISM is simulated as a collisional component,
i.e. as discrete particles which possess a mass and a radius and which
can have inelastic collisions (sticky particles).
Since the ISM is a turbulent and fractal medium (see e.g. Elmegreen \& Falgarone 1996),
it is neither continuous nor discrete. The volume filling factor of the warm and cold phases
is smaller than one. The warm neutral and ionized gas fill about $30-50\%$ of the volume,
whereas cold neutral gas has a volume filling factor smaller than 10\% (Boulares \& Cox 1990). 
It is not clear how this fraction changes, when an external 
pressure is applied. In contrast to smoothed particles hydrodynamics (SPH), which is a 
quasi continuous approach and where the particles cannot penetrate each other, our approach 
allows a finite penetration length, which is given by the mass-radius relation of the particles.
Both methods have their advantages and their limits.
The advantage of our approach is that ram pressure can be included easily as an additional
acceleration on particles that are not protected by other particles (see Vollmer et al. 2001).
In this way we avoid the problem of treating the huge density contrast between the 
ICM ($n \sim 10^{-4}$~cm$^{-3}$) and the ISM ($n > 1$~cm$^{-3}$) of the galaxy.

The 20\,000 particles of the collisional component represent gas cloud complexes which are 
evolving in the gravitational potential of the galaxy.
The total assumed gas mass is $M_{\rm gas}^{\rm tot}=8.9\,10^{9}$~M$_{\odot}$,
which corresponds to the total neutral gas mass before stripping.
Following Vollmer \& Huchtmeier (2007) we assume an initial gas disk which is
truncated at approximately the optical radius ($\sim 17$~kpc).
To each particle a radius is attributed depending on its mass. 
During the disk evolution the particles can have inelastic collisions, 
the outcome of which (coalescence, mass exchange, or fragmentation) 
is simplified following Wiegel (1994). 
This results in an effective gas viscosity in the disk. 

As the galaxy moves through the ICM, its gas clouds are accelerated by
ram pressure. Within the galaxy's inertial system its clouds
are exposed to a wind coming from a direction opposite to that of the galaxy's 
motion through a static ICM. 
The temporal ram pressure profile has the form of a Lorentzian,
which is a good approximation for galaxies on highly eccentric orbits within the
Virgo cluster (Vollmer et al. 2001).
The effect of ram pressure on the clouds is simulated by an additional
force on the clouds in the wind direction. Only clouds which
are not protected by other clouds against the wind are affected.

The particle trajectories are integrated using an adaptive timestep for
each particle. This method is described in Springel et al. (2001).
The following criterion for an individual timestep is applied:
\begin{equation}
\Delta t_{\rm i} = \frac{20~{\rm km\,s}^{-1}}{a_{\rm i}}\ ,
\end{equation}
where $a_{i}$ is the acceleration of the particle i.
The minimum value of $t_{\rm i}$ defines the global timestep used 
for the Burlisch--Stoer integrator that integrates the collisional
component.

\subsection{Search for the best fit model \label{sec:bestfit}}

\begin{figure} 
	\resizebox{\hsize}{!}{\includegraphics{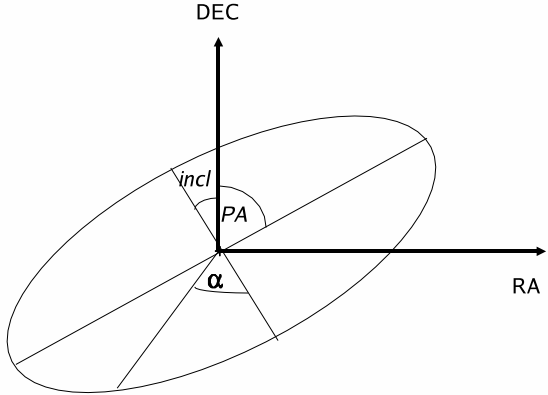}}
	\resizebox{\hsize}{!}{\includegraphics{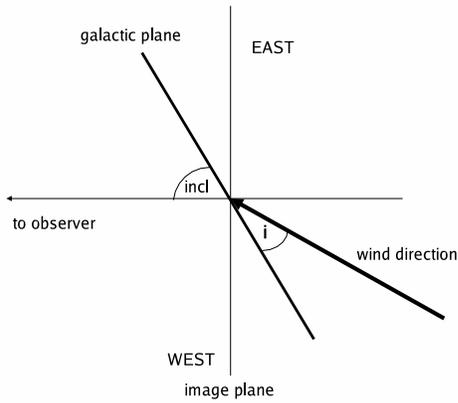}}
	\caption{Galaxy geometry. Upper panel: image plane. $PA$: position angle; 
	  incl: inclination of the galactic
	  disk with respect to the plane of the sky; $\alpha$: azimuthal viewing angle
	  within the galaxy plane defined by $PA$ and $incl$.
	  Lower panel: perpendicular view from the top. $i$: disk-wind angle.
	} \label{fig:dessin1}
\end{figure}

In this section we will constrain the parameters of the ram pressure stripping event
which are (i) the peak ram pressure, (ii) the temporal ram pressure profile,
(iii) time since peak ram pressure, (iv) the disk-wind angle between the galaxy's 
disk and the intracluster medium wind direction, and (v) the azimuthal viewing angle for 
the observed inclination and position angles (Fig.~\ref{fig:dessin1}).
These parameters are related to the observed quantities which are (1) the position angle $PA$,
(2) the inclination angle of the galactic disk $incl$, (3) the line-of-sight velocity
of the galaxy with respect to the cluster mean, and (4) the projected ICM wind direction.
The position angle and inclination of NGC~4501 define the galaxy plane in three dimensional space.
The model galaxy can then be rotated within this plane by the azimuthal viewing angle (see below). 
The three dimensional model wind direction, the line-of-sight velocity of the galaxy, and
the projected ICM wind direction are thus functions of the azimuthal viewing angle.

\begin{figure} 
	\resizebox{\hsize}{!}{\includegraphics{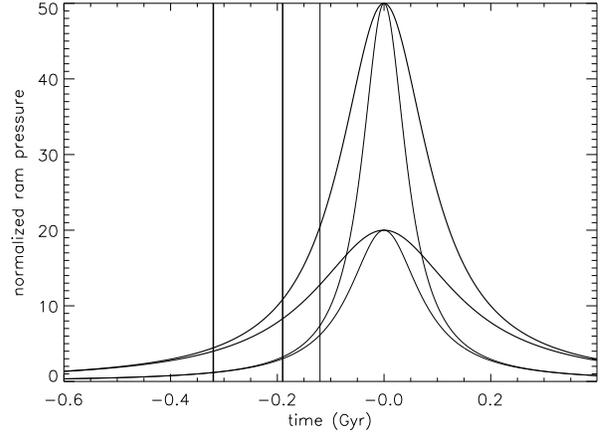}}
	\caption{The adopted temporal ram pressure profiles 
	  for the low ram pressure (LRP) and the high ram pressure (HRP) model.
	  Thick lines: ram pressure profiles with doubled widths (LRP1/2, HRP1/2).
	  The vertical lines show the adopted timesteps for the
	  comparison between model and observations.
	} \label{fig:rpsprofiles}
\end{figure}

We assume a highly eccentric galaxy orbit within the Virgo cluster leading to
a temporal ram pressure profile of the form: 
\begin{equation}
p_{\rm ram}=p_{\rm max} \frac{t_{\rm HW}^{2}}{t^{2}+t_{\rm HW}^{2}}\ ,
\label{eq:rps}
\end{equation}
where $t_{\rm HW}$ is the width of the profile (Vollmer et al. 2001).
We define $t=0$~Myr as the time when ram pressure is maximum.
As shown in Vollmer et al. (2001) the width of the profile depends
on the ram pressure maximum. For a high maximum ram pressure the galaxy
has to come very close to the cluster center and consequently its
velocity is higher while it crosses the high ICM density region in the
cluster core.
We use two typical profiles for strong and
moderately strong ram pressure stripping (Fig.~\ref{fig:rpsprofiles},
see Vollmer et al. 2001):
\begin{enumerate}
\item
low ram pressure model (LRP):\\
$p_{\rm max}=2000$~cm$^{-3}$(km\,s$^{-1}$)$^{2}$, $t_{\rm HW}=80$~Myr and
\item
high ram pressure model (HRP):\\
$p_{\rm max}=5000$~cm$^{-3}$(km\,s$^{-1}$)$^{2}$, $t_{\rm HW}=50$~Myr.
\end{enumerate}
It turned out that, in both simulations, NGC~4501 encounters peak ram pressure
within $120$~Myr. Since this is not consistent with its location within the cluster 
assuming classical galaxy orbits in a static smooth intracluster medium (Vollmer et al. 2001),
we made a second set of simulations for which we doubled the widths of the temporal ram pressure 
profiles (Fig.~\ref{fig:rpsprofiles}): 
\begin{enumerate}
\item
low ram pressure model (LRP1/2):\\
$p_{\rm max}=2000$~cm$^{-3}$(km\,s$^{-1}$)$^{2}$, $t_{\rm HW}=160$~Myr and
\item
high ram pressure model (HRP1/2):\\
$p_{\rm max}=5000$~cm$^{-3}$(km\,s$^{-1}$)$^{2}$, $t_{\rm HW}=100$~Myr.
\end{enumerate}

\begin{figure} 
	\resizebox{\hsize}{!}{\includegraphics{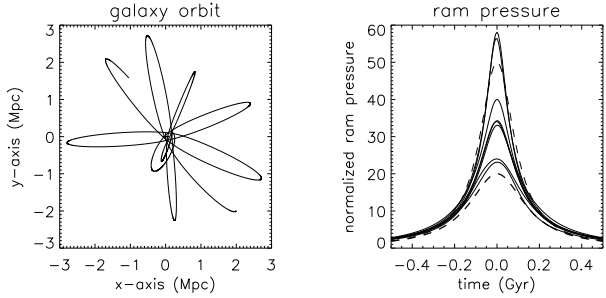}}
	\caption{Left panel: galaxy orbit in the Virgo cluster leading to temporal ram
	  pressure profiles with a larger width. Right panel:
	  resulting ram pressure profiles (solid lines) and
	  LRP1/2, HRP1/2 ram pressure profiles (dashed lines).
	} \label{fig:n4501_orbit_rps}
\end{figure} 
In these simulations the galaxy does not go as close to the cluster core 
(Fig.~\ref{fig:n4501_orbit_rps}) as in the LRP and HRP models.
The minimum distance to the cluster center is about $10$\,\% larger than that
of the orbits shown in Vollmer et al. (2001). 
The minimum distance of the LRP1/2 is about $0.25$~Mpc, that of the HRP1/2 model is about $0.1$~Mpc.
The larger width of the ram pressure profile is mainly due to a broader 
intracluster medium distribution compared to the LRP and HRP simulations.
 
\begin{figure*} 
	\resizebox{14cm}{!}{\includegraphics{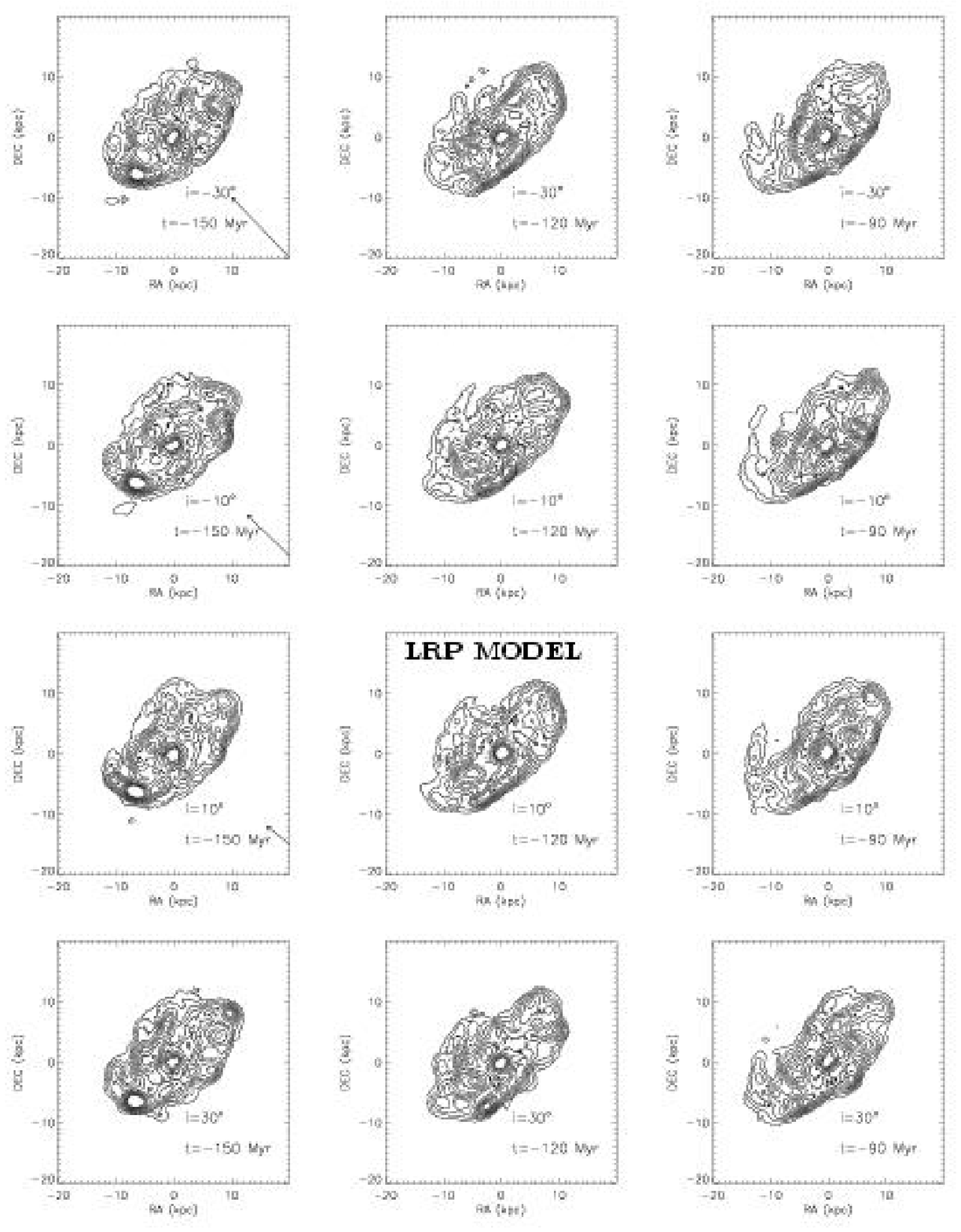}}
	\caption{Model snapshots of the gas column density 
	 for different simulations of the low ram pressure (LRP) model.
	 Contours are (1, 2, 3, 4, 5, 6, 7, 8, 9, 10)$\times 4\,10^{20}$~cm$^{-2}$.
	 The ram pressure profile for all simulations isq
	 given by Eq.~\ref{eq:rps}. The disk-wind angle $i$
	 between the orbital and the galaxy's disk plane and
	 the timestep of the snapshot are varied.
	 Left column: $t=-150$~Myr. Middle column: $t=-100$~Myr.
	 Right column: $t=-50$~Myr. Upper row: $i=-30^{\circ}$.
	 Middle rows: $i=-10^{\circ}$ and $i=10^{\circ}$. Lower row: $i=30^{\circ}$.
	 The arrow represents the projected ICM wind direction. 
	 The absence of the arrow for $i=30^{\circ}$ is due to a
	 zero projected galaxy velocity.
	} \label{fig:n4501_p20}
\end{figure*}

\begin{figure*} 
	\resizebox{14cm}{!}{\includegraphics{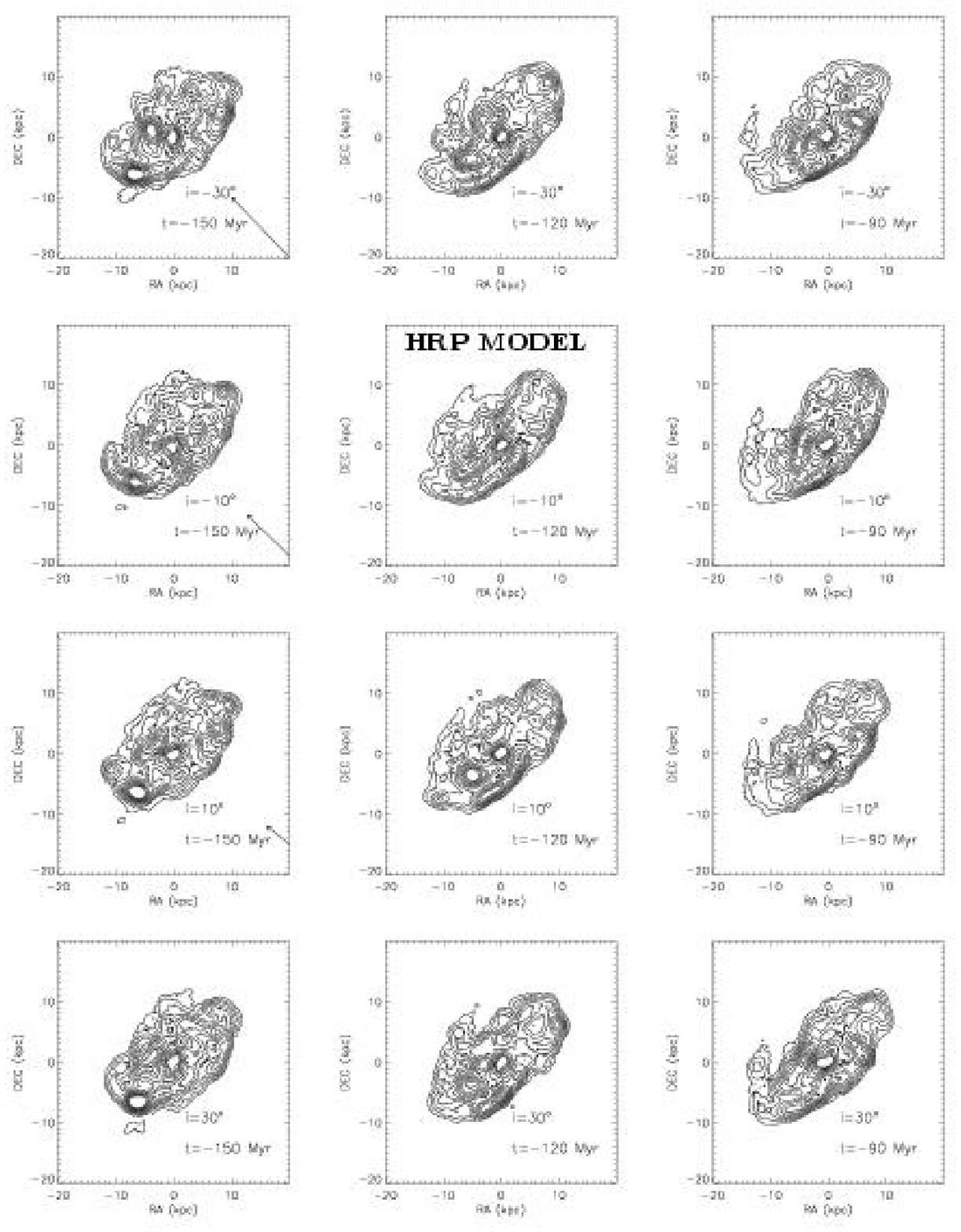}}
	\caption{Model snapshots of the gas column density
	 for different simulations of the high ram pressure (HRP) model.
	 Contours are the same as in Fig.~\ref{fig:n4501_p20}.
	 The ram pressure profile for all simulations is
	 given by Eq.~\ref{eq:rps}. The disk-wind angle $i$
	 between the orbital and the galaxy's disk plane and
	 the timestep of the snapshot are varied.
	 Left column: $t=-150$~Myr. Middle column: $t=-100$~Myr.
	 Right column: $t=-50$~Myr. Upper row: $i=-30^{\circ}$.
	 Middle rows: $i=-10^{\circ}$ and $i=10^{\circ}$. Lower row: $i=30^{\circ}$.
	 The absence of the arrow for $i=30^{\circ}$ is due to a
	 zero projected galaxy velocity.
	} \label{fig:n4501_p50}
\end{figure*}

\begin{figure*} 
	\resizebox{14cm}{!}{\includegraphics{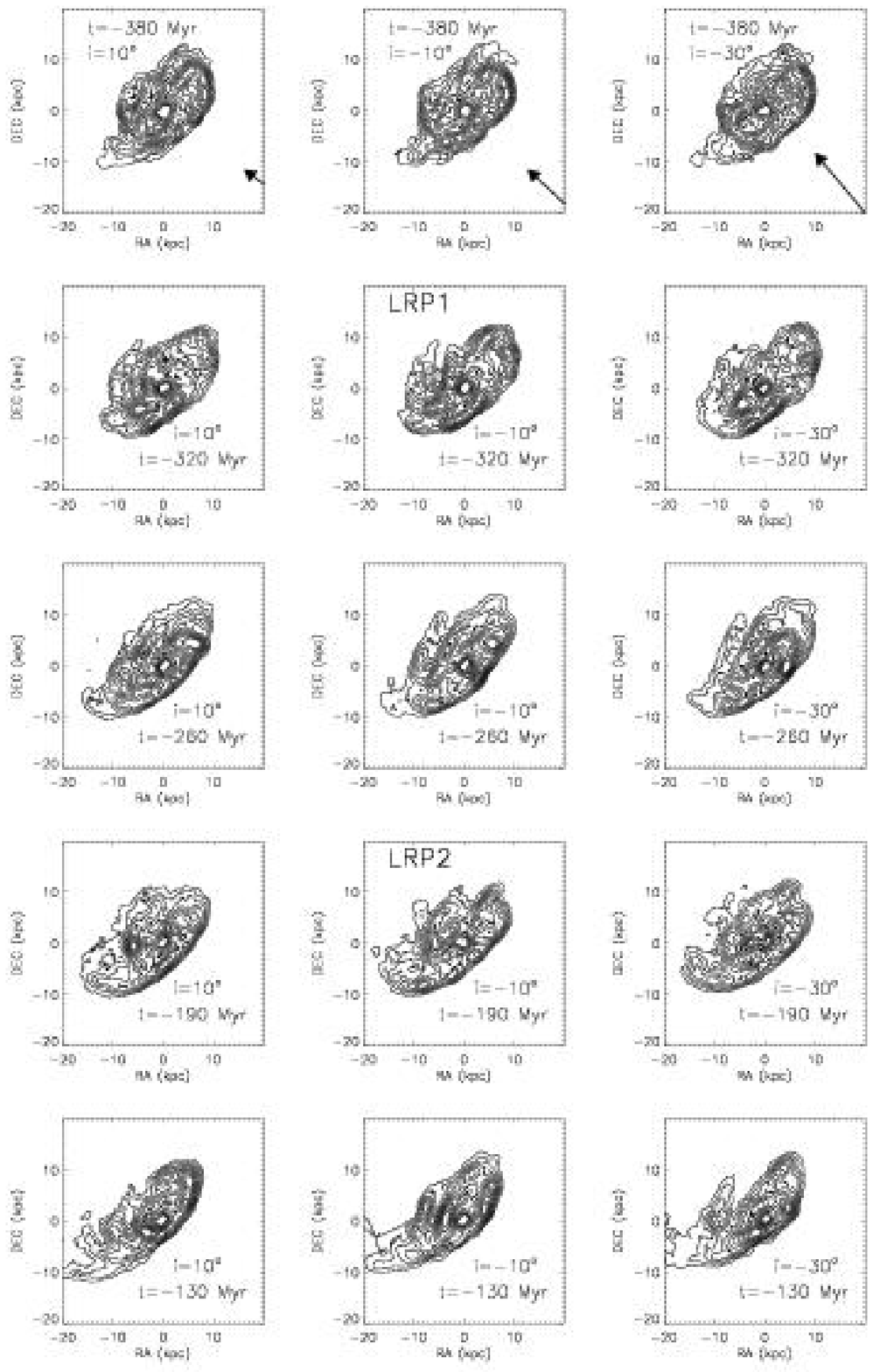}}
	\caption{Model snapshots of the gas column density
	 for different simulations of the low ram pressure (LRP1/2) model
	 with the large temporal ram pressure profile.
	 Contours are the same as in Fig.~\ref{fig:n4501_p20}.
	 The ram pressure profile for all simulations is
	 given by Eq.~\ref{eq:rps}. The disk-wind angle $i$
	 between the orbital and the galaxy's disk plane and
	 the timestep of the snapshot are varied.
	} \label{fig:n4501_plot_p20}
\end{figure*}

\begin{figure*} 
	\resizebox{14cm}{!}{\includegraphics{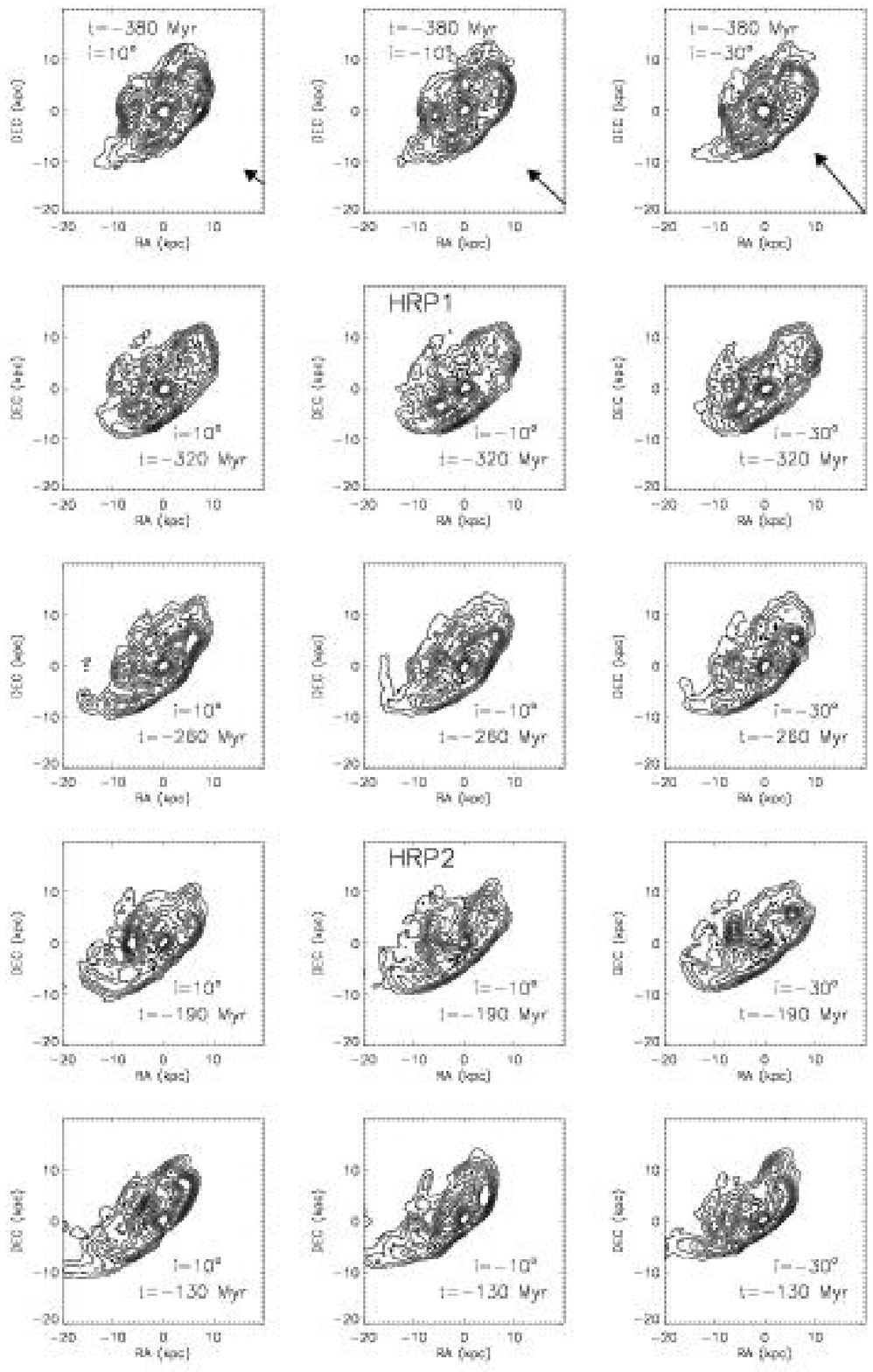}}
	\caption{Model snapshots of the gas column density
	 for different simulations of the high ram pressure (HRP1/2) model
	 with the large temporal ram pressure profile.
	 Contours are the same as in Fig.~\ref{fig:n4501_p20}.
	 The ram pressure profile for all simulations is
	 given by Eq.~\ref{eq:rps}. The disk-wind angle $i$
	 between the orbital and the galaxy's disk plane and
	 the timestep of the snapshot are varied.
	} \label{fig:n4501_plot_p50}
\end{figure*}

With these 4 simulations we aim to reproduce the main characteristics of our H{\sc i}
observations:
\begin{itemize}
\item
the low surface-density gas located northeast of the galaxy center,
\item
the region of high surface-density gas at the opposite windward side,
\item
the asymmetry along the major axis with a more extended emission to the 
southeast,
\item
the rising rotation curve to the northwest,
\item
the southwestern ridge of polarized radio-continuum emission. 
\end{itemize}
The first snapshot selection is mainly based on the H{\sc i} morphology.
For the LRP and HRP simulations we identify one best fit model
(LRP: Fig.~\ref{fig:n4501_p20}, HRP: Fig.~\ref{fig:n4501_p50}),
for the LRP1/2 and HRP1/2 simulations two best fit models (LRP1 and LRP2:
Fig.~\ref{fig:n4501_plot_p20}; HRP1 and HRP2: Fig.~\ref{fig:n4501_plot_p50}) are selected.
In a second step (Sect.~\ref{sec:comparison}), gas kinematics and the polarized radio contiuum
emission distribution are taken into account. Table~\ref{tab:summary}
summarizes our findings.
\begin{table*}
      \caption{Features which are reproduced by the different models.}
         \label{tab:summary}
      \[
         \begin{array}{lcccccc}
           \hline
           \noalign{\smallskip}
           {\rm model} & {\rm LRP} & {\rm HRP} & {\rm LRP1} & {\rm LRP2} & {\rm HRP1} & {\rm HRP2} \\
           \hline
           {\rm NE\ low\ surface\ density\ gas} & + & + & + & + & + & + \\
	   {\rm SW\ high\ surface\ density\ gas\ ridge}& + & + & + & + & + & + \\
	   {\rm H{\sc I}\ asymmetry\ along\ major\ axis}& - & - & - & + & - & + \\
	   {\rm NW\ rising\ rotation\ curve} & + & + & + & + & + & + \\
	   {\rm SW\ ridge\ of\ polarized\ radio\ emission} & + & + & + & + & + & + \\
	   {\rm symmetric\ velocity\ dispersion\ distribution} & + & + & + & - & + & - \\
	   {\rm consistency\ with\ galaxy\ orbit} & - & - & + & ? & + & ? \\
           \noalign{\smallskip}
        \hline
        \end{array}
      \]
\end{table*}

\begin{figure} 
	\resizebox{\hsize}{!}{\includegraphics{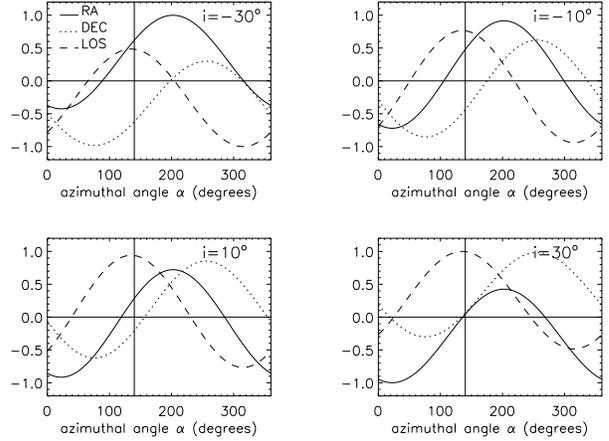}}
	\caption{Components of the 3D direction of the galaxy's velocity as 
	  a function of the azimuthal viewing angle $\alpha$.
	  The length of the vector is unity.
	  Solid line: right ascension; dotted line: declination;
	  dashed line: line-of-sight direction for four different
	  disk-wind angle between the galactic disk and the ICM wind
	  $i=-30^{\circ}$, $i=-10^{\circ}$, $i=10^{\circ}$, and $i=30^{\circ}$.
	  The vertical line represents a viewing angle of $140^{\circ}$.
	} \label{fig:winkel}
\end{figure}

Ram pressure efficiency depends on the disk-wind angle $i$
between the galactic disk and the ICM wind direction (Vollmer et al. 2001).
A disk-wind angle of $i < -30^{\circ}$
is not compatible with NGC~4501's positive radial velocity with respect to
the Virgo cluster. Disk-wind angles $i > 30^{\circ}$ lead to projected
wind directions to the northeast for almost all azimuthal viewing angles. 
However, the H{\sc i} and polarized emission observations 
indicate that the projected wind direction has an important component to the west.
Therefore, we do not consider stripping with $|i| > 30^{\circ}$.
For each model (LRP and HRP) we made 4 simulations with 4 different 
disk-wind angles between the galaxy's disk and the ICM wind direction:
(i) $i=-30^{\circ}$, (ii) $i=-10^{\circ}$, (iii) $i=10^{\circ}$, and (iv) $i=30^{\circ}$.
A disk-wind of $i=0^{\circ}$ means that the galactic disk is
parallel to the ICM wind direction.
The azimuthal viewing angle, which is the angle in the galactic plane
defined by the galaxy's inclination and position angles, is chosen
in a way to fit the observed H{\sc i} distribution, i.e. the position
of the extended northeastern low surface-density gas and to reproduce
the positive line-of-sight component of the wind direction (the galaxy is moving away 
from the observer).
In Figs.~\ref{fig:winkel} we show the components of the 3D ICM wind direction.
For the chosen azimuthal viewing angle $\alpha=140^{\circ}$ the line-of-sight
velocity is positive and represents between 50\% and 100\% 
of the total velocity with respect to the cluster mean. 

In Figs.~\ref{fig:n4501_p20} and \ref{fig:n4501_p50} we show snapshots of the
LRP and HRP model at 3 different timesteps $-150$~Myr, $-100$~Myr, and $-50$~Myr, i.e.
before the ram pressure maximum.
All model gas distributions at $t=-150$~Myr have symmetric outer contours and
a local surface-density maximum in the southeast, which is due to the internal
spiral structure. At this timestep ram pressure is not strong enough to
affect the gas distribution of the disk significantly. However, gas of lower
surface density is already displaced for $t < -150$~Myr.
At $t=-120$~Myr
the southwestern edge of the gas disk is now compressed and shows a sharp edge. 
On the opposite side, i.e. in the
northeast, gas is pushed to larger galactic radii. Since the gas is
still rotating there, it forms a kind of detached low surface spiral arm. 
This arm is located somewhat closer to the disk for $i=30^{\circ}$
than for $i=-30^{\circ}$, because it is located above the disk which
makes its projected distance smaller.
At $t=-90$~Myr the northeastern low surface arm has moved to the southeast 
and its projected distance from the main disk has increased.
The LRP and HRP models are indistinguishable at $t=-150$~Myr. 
At $t=-120$~Myr the overall H{\sc i} morphologies of the LRP and HRP model
snapshots are similar and it is mainly the structure of the surface density distribution 
of the northeastern arm and the gas disk which change from one
model to another. The latter statement also applies for the snapshots at $t=-90$~Myr.
Based on these snapshots, only a simulation snapshot at $t=-120$~Myr 
reproduces the observed southwestern compression region and the
northeastern low surface-density arm. In all model snapshots at $t=-120$~Myr
we observe a region of high surface density $\sim 5$~kpc southwest of the
galaxy center, which is not present in the H{\sc i} data. Since this is
due to the internal structure of the model galaxy, which (i) is not affected by
ram pressure because of its location in the inner galactic disk and (ii) depends on the
initial conditions, we do not take it into account for the comparison between 
models and observations. Based on a detailed 
comparison of the model gas distribution, including the southwestern 
overdensities and the northeastern low surface-density arm, we chose
two final snapshots with $i=10^{\circ}$ for the LRP model and
$i=-10^{\circ}$ for the HRP model labeled in Figs.~\ref{fig:n4501_p20} and 
\ref{fig:n4501_p50}.
 
Both simulations yield a timescale of 120~Myr before peak stripping.
The galaxy is moving to the southwest with a velocity component in the
sky-plane of $\sim 1000$~km\,s$^{-1}$. Until peak ram pressure it will
cover a distance of 0.13~Mpc. Its actual distance from the cluster center
at peak ram pressure will be greater than 0.5~Mpc. However, typical orbits
associated to the ram pressure profiles LRP and HRP have impact parameters 
of $0.1 - 0.2$~Mpc (Vollmer et al. 2001).
To reconcile this difference, we doubled the widths of the
temporal ram pressure profiles (LRP1/2 and HRP1/2 simulations; Fig.~\ref{fig:rpsprofiles}).
This is still within the uncertainty of the intracluster medium density
distribution given by Schindler et al. (1999). The impact parameters
for orbits associated with these ram pressure profiles are
$0.2 - 0.4$~Mpc. Since $i=30^{\circ}$ lead
to negligible velocity components in the plane of the sky for the adopted 
azimuthal viewing angle (Fig.~\ref{fig:winkel}), we do not consider this possibility anymore.

In Figs.~\ref{fig:n4501_plot_p20} and \ref{fig:n4501_plot_p50} we show snapshots of the
LRP1/2 and HRP1/2 model at 5 different timesteps $-380$~Myr, $-320$~Myr, $-260$~Myr,
$-190$~Myr, and $-130$~Myr, i.e. before ram pressure maximum.
The evolution of the LRP1/2 and HRP1/2 models for a given disk-wind angle 
between the disk and the ram pressure wind are qualitatively similar. 
For timesteps $t < -150$~Myr the H{\sc i} distribution of the different disk-wind
angles are also similar. At $t=-380$~Myr there
is more high surface density in the northwestern part of the galaxy, whereas we observe
a low surface-density arm in the southeast. At $t=-320$~Myr the disk emission shows
is approximately symmetric along the major axis and a low surface-density arm 
is formed east of the galaxy center. This snapshot corresponds to the LRP and
HRP snapshots at $t=-120$~Myr (Figs.~\ref{fig:n4501_p20} and \ref{fig:n4501_p50}), because
the ram pressure induced momentum transfer at the two respective timesteps is about the same
(Fig.~\ref{fig:rpsprofiles}). At $t=-260$~Myr this arm is located closer to the
galaxy center and again a southeastern low surface-density arm is formed.
A pronounced asymmetry along the minor axis is observed at $t=-190$~Myr with
more emission in the compressed western part of the disk. In addition, we find a less pronounced
asymmetry along the major axis with more emission in the southeastern part.
At $t=-130$~Myr the gas in the western part of the disk is pushed to smaller radii and
we observe a low surface-density arm in the southeast of the galaxy center.
Based on this set of simulations, 4 snapshots with $i=-10^{\circ}$ reproduce our H{\sc i} observations:
LRP1 - $t=-320$~Myr, LRP2 - $t=-190$~Myr, HRP1 - $t=-320$~Myr, and HRP2 - $t=-190$~Myr.
All 4 snapshots show a low surface-density gas located northeast of the galaxy center and
an region of high surface-density gas at the opposite windward side. Only the snapshots at
$t=-190$~Myr show an asymmetry along the major axis with a more extended emission to the 
southeast.

\begin{figure}
	\resizebox{7cm}{!}{\includegraphics{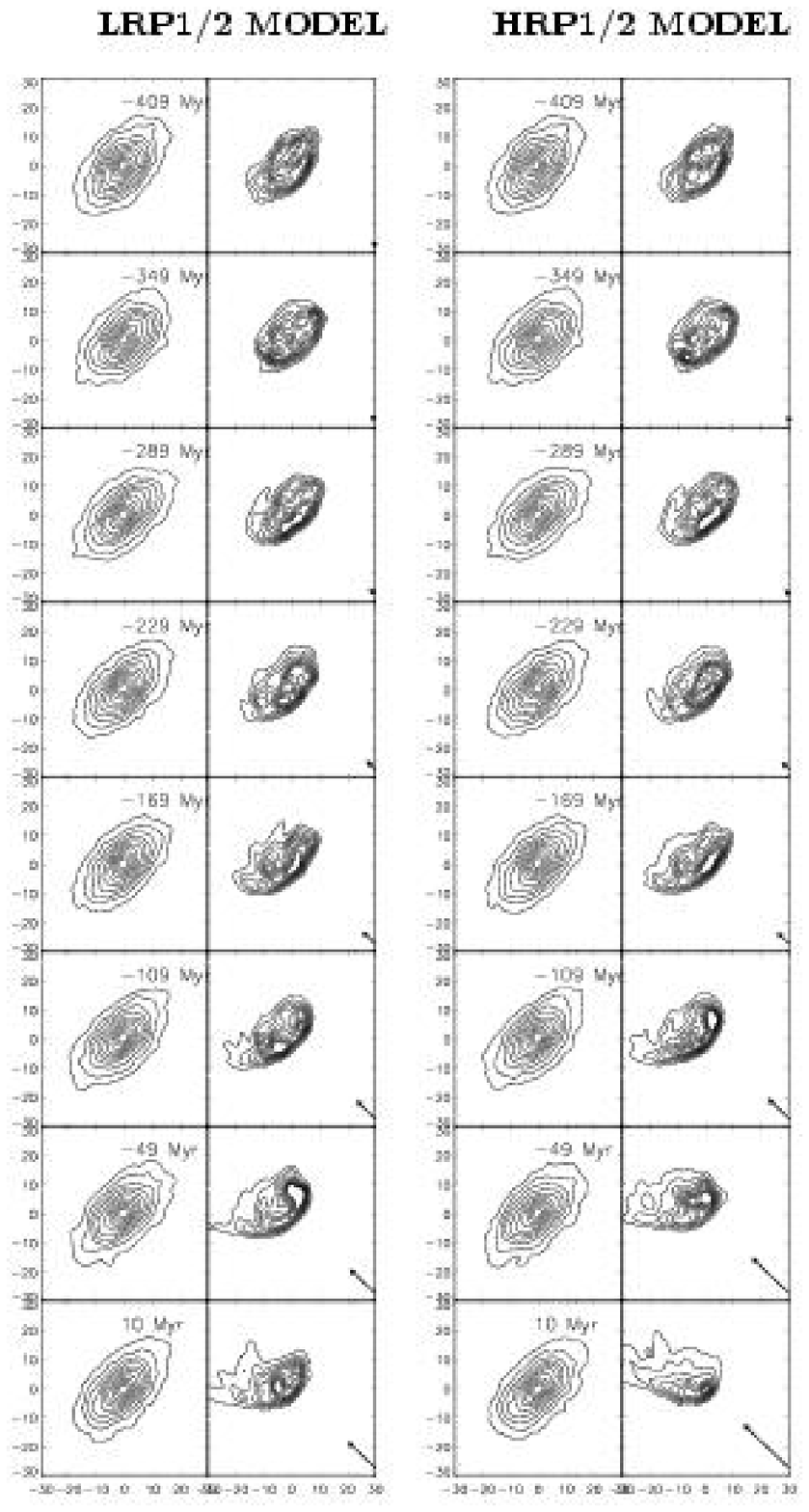}}
	\caption{Evolution of the model stellar (1st column)
	and gas disk (2nd column).  
	The contours of the stellar distribution are logarithmic,  
	those of the gas distribution are the same as in Fig.~\ref{fig:n4501_p20}.
	The arrow indicates the direction of ram pressure, i.e. it is opposite to
	the galaxy's velocity vector, and its size is proportional
	to $\rho v_{\rm gal}^{2}$. Maximum ram pressure occurs at
	$t=0$~Myr. The timestep of each snapshot is marked in each panel
	showing the stellar disk.
	} \label{fig:evolution}
\end{figure}

The time evolutions of these simulations are shown in Fig.~\ref{fig:evolution}.
Since ram pressure only affects the interstellar medium of the galaxy,
the stellar disk stays symmetric during the whole simulation.
The sky-projected wind direction is northeastward for the LRP1/2 and HRP1/2 model.
As expected, the stripping radius of the HRP1/2 model is smaller
than that of the LRP1/2 model, i.e. the galaxy in the HRP1/2 model
loses more gas than that of the LRP1/2 model.


\subsection{Magnetic field evolution \label{sec:mhdmodel}}

Otmianowska-Mazur \& Vollmer (2003) studied the evolution of the
large-scale magnetic field during a ram pressure stripping event.
They calculated the magnetic field structure by solving the induction equation 
on the velocity fields produced by the dynamical model.
The polarized radio-continuum emission has been calculated by assuming
a Gaussian spatial distribution of relativistic electrons. This procedure
allowed them to study the evolution of the observable polarized radio
continuum emission during a ram pressure stripping event.

We apply the same procedure as Otmianowska-Mazur \& Vollmer (2003)
to a similar ram pressure stripping event (Sect.~\ref{sec:model}).
The Zeus3D code  (Stone \& Norman 1992a and b) is used to solve
the induction equation: 
\begin{equation}
{\partial\vec{B}/\partial t=\hbox{rot}(\vec{v}\times\vec{B})
 -\hbox{rot}(\eta~\hbox{rot}\vec{B})}
\label{eq:inductioneq}
\end{equation}
where $\vec{B}$ is the magnetic induction, $\vec{v}$ is the large-scale
velocity of the gas, and $\eta$ is the coefficient of a turbulent diffusion.
We use a physical diffusion of $\eta = 5 \times 10^{25}$~cm$^{2}$s$^{-1}$
(Elstner et al. 2000). We do not implement any dynamo process. 
The initial magnetic field is purely toroidal with a strength of 10~$\mu$G.

The induction equation is solved on rectangular coordinates ($XYZ$).
The number of grid points used is 215x215x91 along the $X$, $Y$ and $Z$ axis, 
respectively with a grid size of 400~pc, resulting in a size of the modeled box of
86~kpc~$\times$~86~kpc~$\times$~36~kpc. Since the N-body code is discrete
whereas the MHD code is using a grid, we have to interpolate the
discrete velocities on the grid. This is done using a method
known as ``Kriging'' with a density-dependent smoothing length
(Isaaks \& Srivastava 1989). It turned out that we
had to use a large smoothing length  to suppress the noise 
in the velocity field of the outer disk, which is due to a small local particle density.
In this way we avoid numerical artifacts in the velocity field
at the outer disk (see Soida et al. 2006).
As the spline interpolation used in Otmianowska-Mazur \& Vollmer (2003),
this has the consequence that the velocity field at the edge of the gas
distribution is more extended than the gas distribution itself.
Since there are gradients in this velocity field due to differential rotation, induction leads to a 
magnetic field which extends beyond the edge of the gas distribution.
This affects the polarized emission beyond the gas distribution, but
not inside, which is what we are interested in.

\begin{figure*} 
	\resizebox{\hsize}{!}{\includegraphics{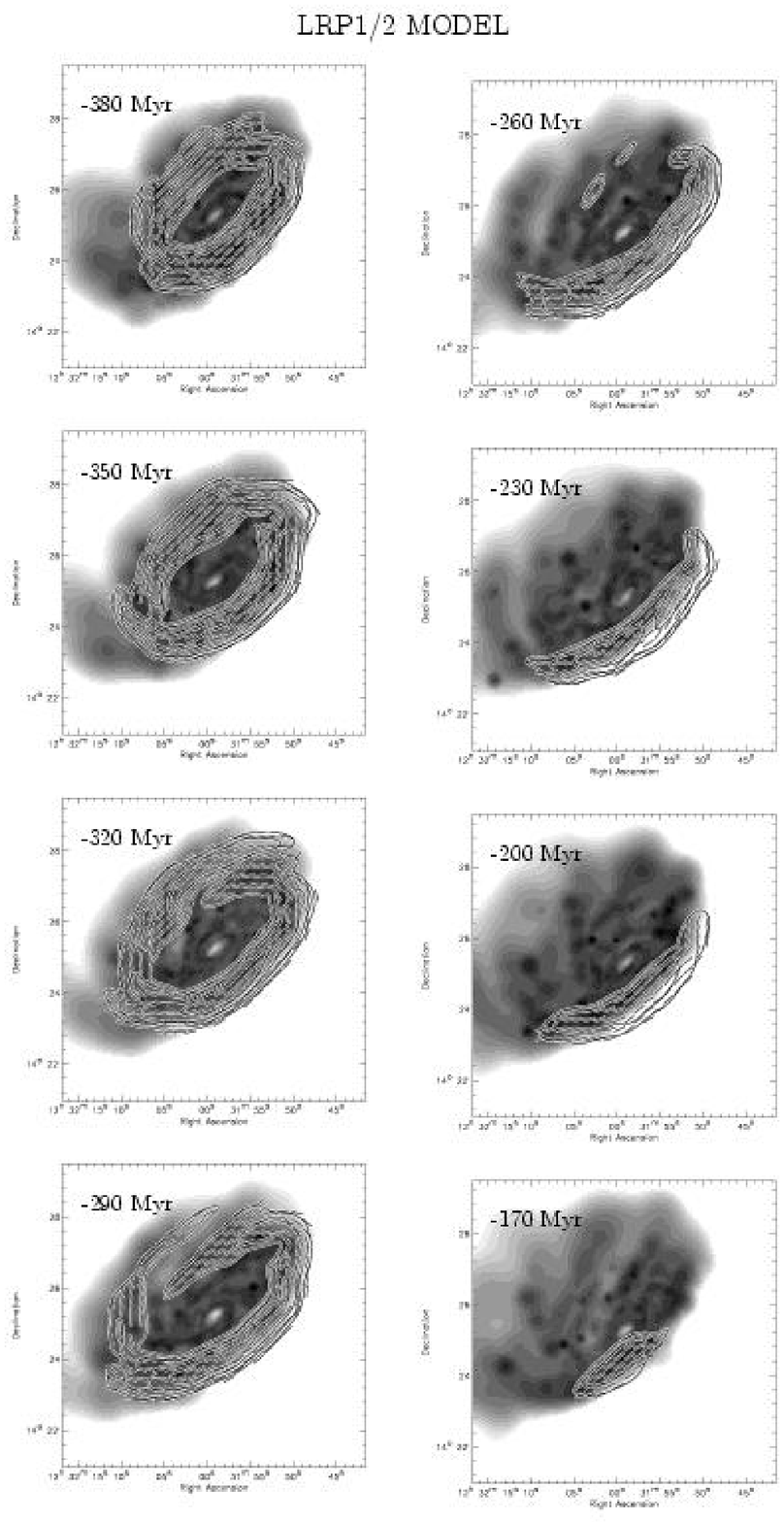}\includegraphics{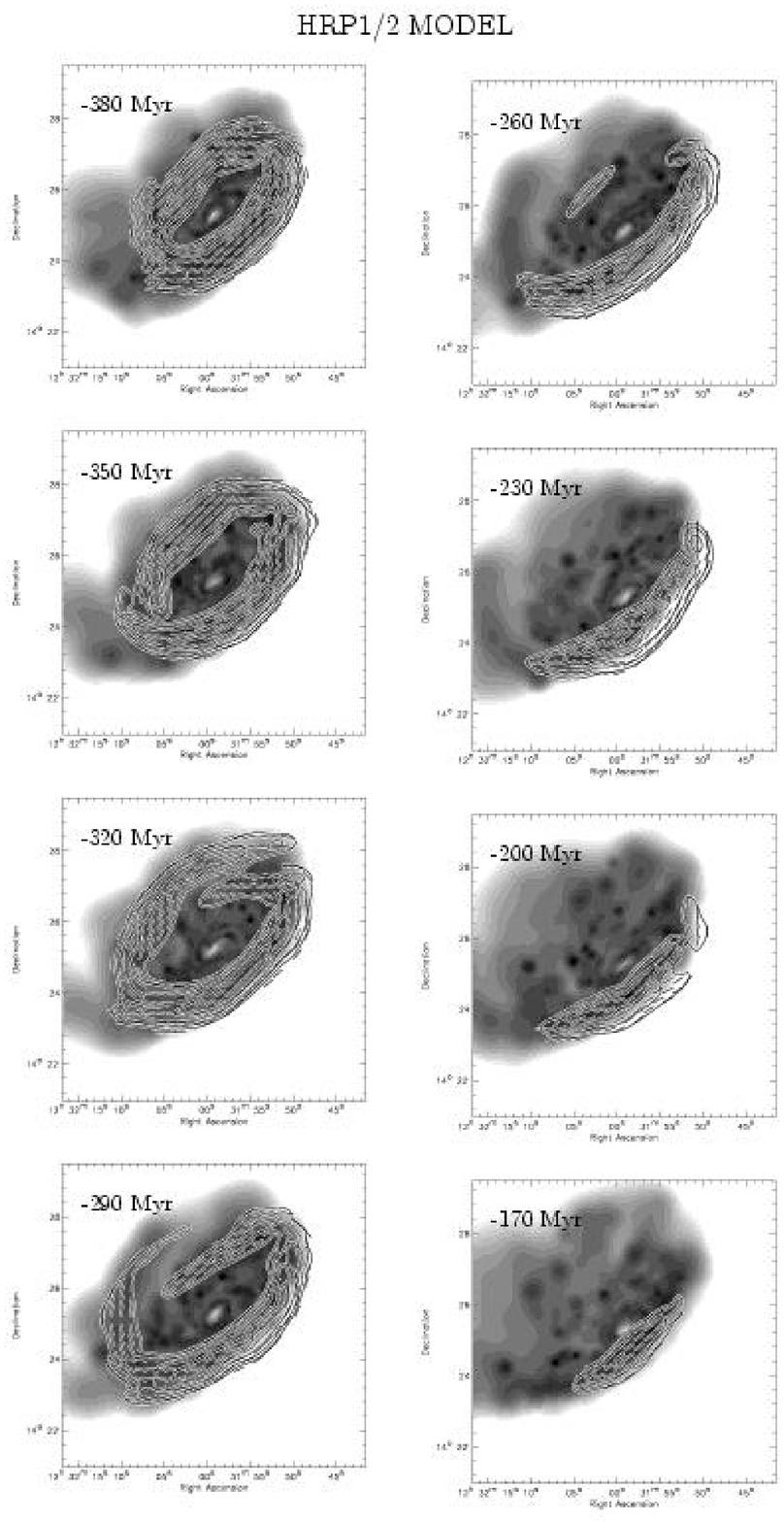}}
	\caption{Evolution of the polarized radio-continuum emission. Time steps
	  are indicated in the upper left corner of each panel. The disk is seen with the position
	  and inclination of NGC~4501 (Table~\ref{tab:parameters}) and
	  rotates counter-clockwise. Greyscale: gas surface density in logarithmic scale. 
	  Contours: polarized radio-continuum emission. The magnetic field
	  vectors are superimposed onto the gas surface density. 
	  The assumed beamsize for the polarized radio-continuum emission is $20''$. 
	} \label{fig:pievolution}
\end{figure*}

The evolution of the polarized radio-continuum emission without Faraday rotation
is presented in Fig.~\ref{fig:pievolution}. The timesteps are the same as in
Fig.~\ref{fig:evolution}. The disk rotates counter-clockwise.
The gas surface density, which is smoothed
to a resolution of $\sim 100$~pc, is shown in greyscales,
the polarized radio-continuum emission as contours, and the magnetic field vectors
projected on the plane of the sky as lines.
We assume a Gaussian distribution of relativistic electrons in $R$ and $z$ directions:
$n_{\rm rel}=n_{0}\,\exp{\big(-(r/r_{\rm R})^2\big)}\,\exp{\big(-(z/r_{\rm z})^2\big)}$, 
where $r_{\rm R}=6.3$~kpc and $r_{\rm z}=1.0$~kpc. 
This translates into a FWHM of 5.2~kpc and 0.83~kpc, respectively.
By assuming this smooth distribution we imply no local equipartition between the energy
densities of total cosmic rays and total magnetic field. 
The galaxy moves to the lower right corner, i.e. the ram pressure wind comes
from this direction. Our grid size does not permit us to
resolve the rotation in the inner part of the galaxy. The errors due to discretization
lead to an artificial radial diffusion of the magnetic field
out of the galactic disk. During the evolution without a significant external influence
($-380$~Myr$< t < -290$~Myr) the model distribution of polarized radio-continuum
emission is approximately symmetric. Once ram pressure compresses the gas, an enhanced ridge of
polarized emission appears in the southwest. At $t=-230$~Myr the internal magnetic
field is lost due to the artificial diffusion, and only the compression induced
ridge of polarized emission is visible. The evolution of polarized radio-continuum emission
is very similar in the LRP1/2 and HRP1/2 models.

\section{Comparison between the models and observations \label{sec:comparison}}

In this section we compare the observed properties of the gas, star formation, and
large-scale magnetic field of NGC~4501 with the properties of the simulated galaxies.
Assuming a static intracluster medium, the projected direction of the ram pressure wind
indicates the projected direction of the galaxy's motion within the intracluster medium.
For the relevant simulations HRP1/2 and LRP1/2 with $i=-10^{\circ}$ this direction is southwest
(Figs.~\ref{fig:n4501_plot_p20} and \ref{fig:n4501_plot_p50}).
A time to peak ram pressure of 320~Myr or 190~Myr leads the galaxy to a distance of
$0.4$~Mpc and $0.5$~Mpc from the cluster center at peak ram pressure assuming a linear
orbit with a constant velocity (see Sect.~\ref{sec:discussion}).  Projection effects will increase,
orbital curvature and acceleration will decrease these values. 
An inspection of simulated galaxy orbits (Fig.~\ref{fig:n4501_orbit_rps}) shows that the difference
between the location at peak ram pressure and the extrapolated location assuming a linear
orbit is $\sim 0.1$~Mpc for a timescale of 320~Myr ($\sim 0.03$~Mpc for a timescale of 190~Myr). 
Our adopted minimum distance scale is thus $0.3$-$0.4$~Mpc. This distance is only compatible
with the low ram pressure simulations LRP1/2 (see Sect.~\ref{sec:bestfit}). 
Therefore, in the following we will only compare the LRP1/2 model to our multi-wavelength observations.

\subsection{The gas moment maps \label{sec:compgas}}

\begin{figure*} 
	\resizebox{\hsize}{!}{\includegraphics{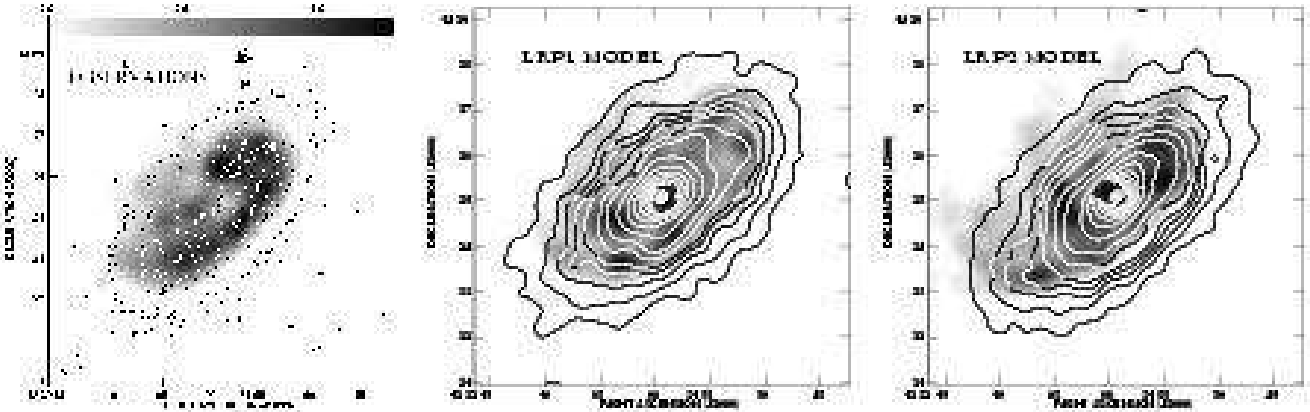}}
	\caption{Comparison between the gas surface density distribution (greyscale)
	  and the stellar surface density distribution (contours).
	  Left panel: B band contours on H{\sc i} surface density distribution (resolution: $30''$).
	  In all figures the greyscales use the full dynamic ranges of the gas distribution.
	}\label{fig:n4501_HI_stars}
\end{figure*}

As described in Sect.~\ref{sec:results} the observed H{\sc i} distribution
follows one isophot contour of the optical image except in the northeast and the north.
Our model snapshots LRP1 and LRP2 show the same qualitative behavior (Fig.~\ref{fig:n4501_HI_stars}).
However, the LRP2 model gas distribution in the west is less extended
than the observed H{\sc i} distribution, i.e. the model gas is pushed to smaller galactic
radii than it is observed. On the other hand, the observed asymmetry along the major axis
is only reproduced by the LRP2 model.

\begin{figure*} 
	\resizebox{\hsize}{!}{\includegraphics{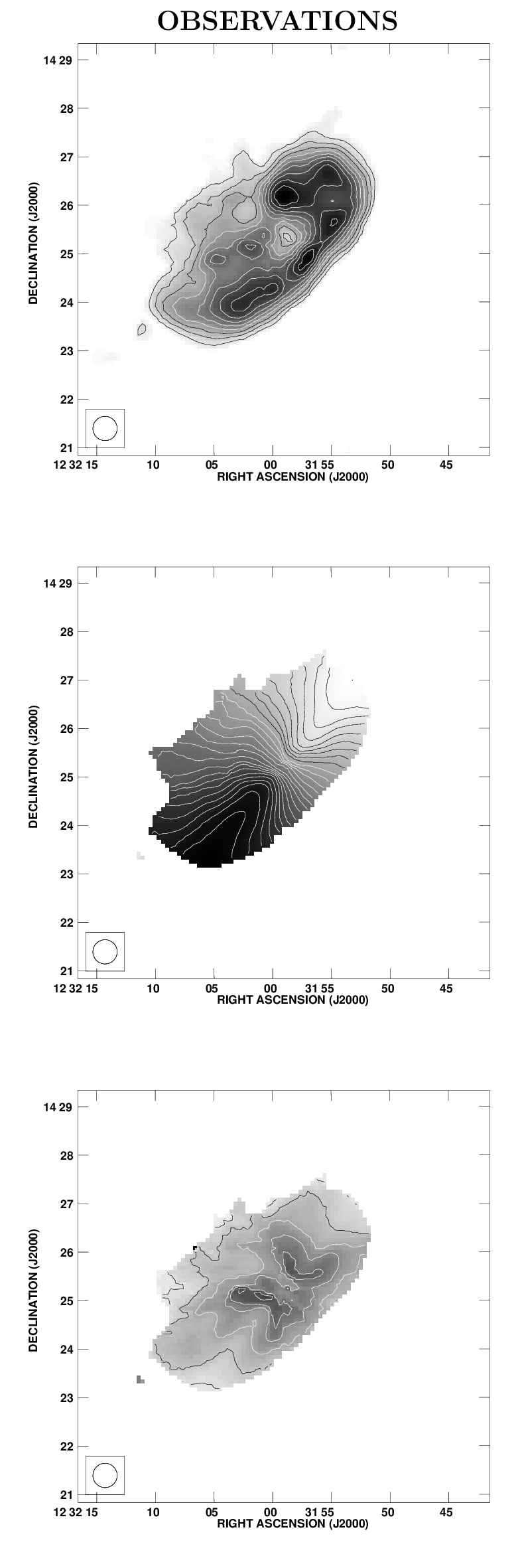}\includegraphics{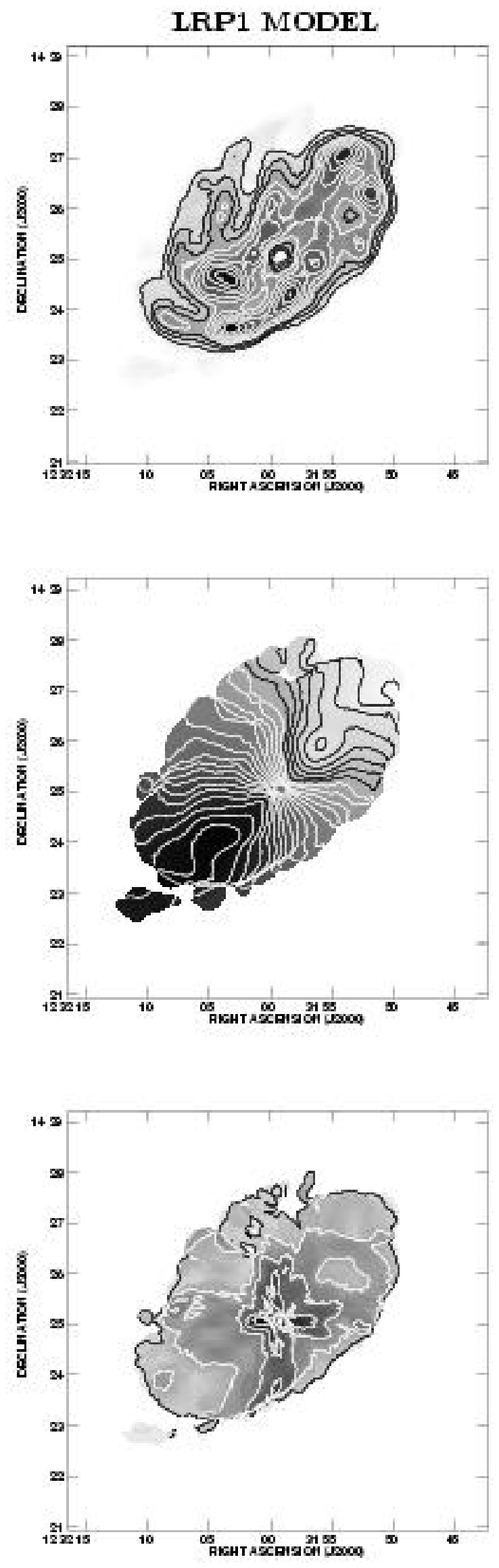}\includegraphics{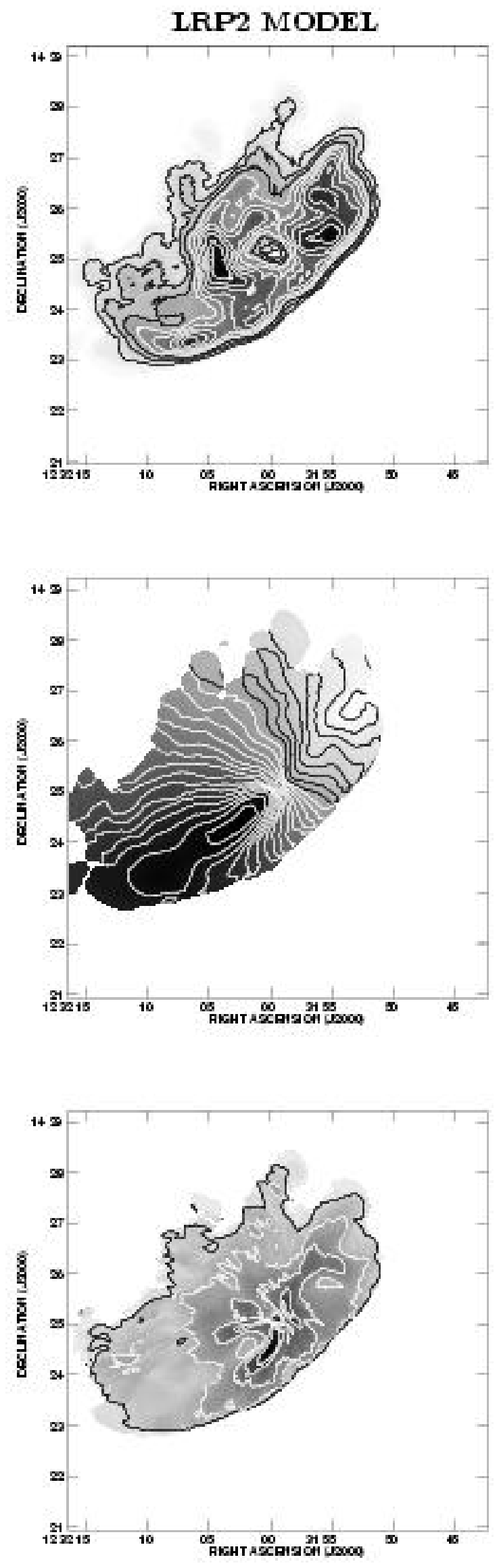}}
	\caption{Comparison between the observed H{\sc i} and simulated gas moment maps.
	  Upper row: gas distribution; middle row: velocity field; lower row: velocity dispersion. 
	} \label{fig:momcomparison}
\end{figure*}

Fig.~\ref{fig:momcomparison} compares the gas distribution, velocity field, and the distribution
of the velocity dispersion of our H{\sc i} observations with our simulations.
Both models reproduce qualitatively the overall observed H{\sc i} surface density distribution
of the outer gas disk. Both models show an overdensity in
the compressed southwestern region of the disk. 
Both models also show an extended low surface-density region in northeast, as it is observed.
There are, however, quantitative differences:
the northeastern low surface-density arm of the LRP1 model is (i) located more to the southeast and (ii) 
more detached from the disk,
whereas that of the LRP2 model is (i) located more to the northwest and (ii) closer to the galaxy center than it is observed.
The LRP2 gas distribution shows a local maximum $\sim 1'$ southeast of the galaxy center.
It is not excluded that a significant amount of this gas might be in molecular form. 

\begin{figure} 
	\resizebox{8cm}{!}{\includegraphics{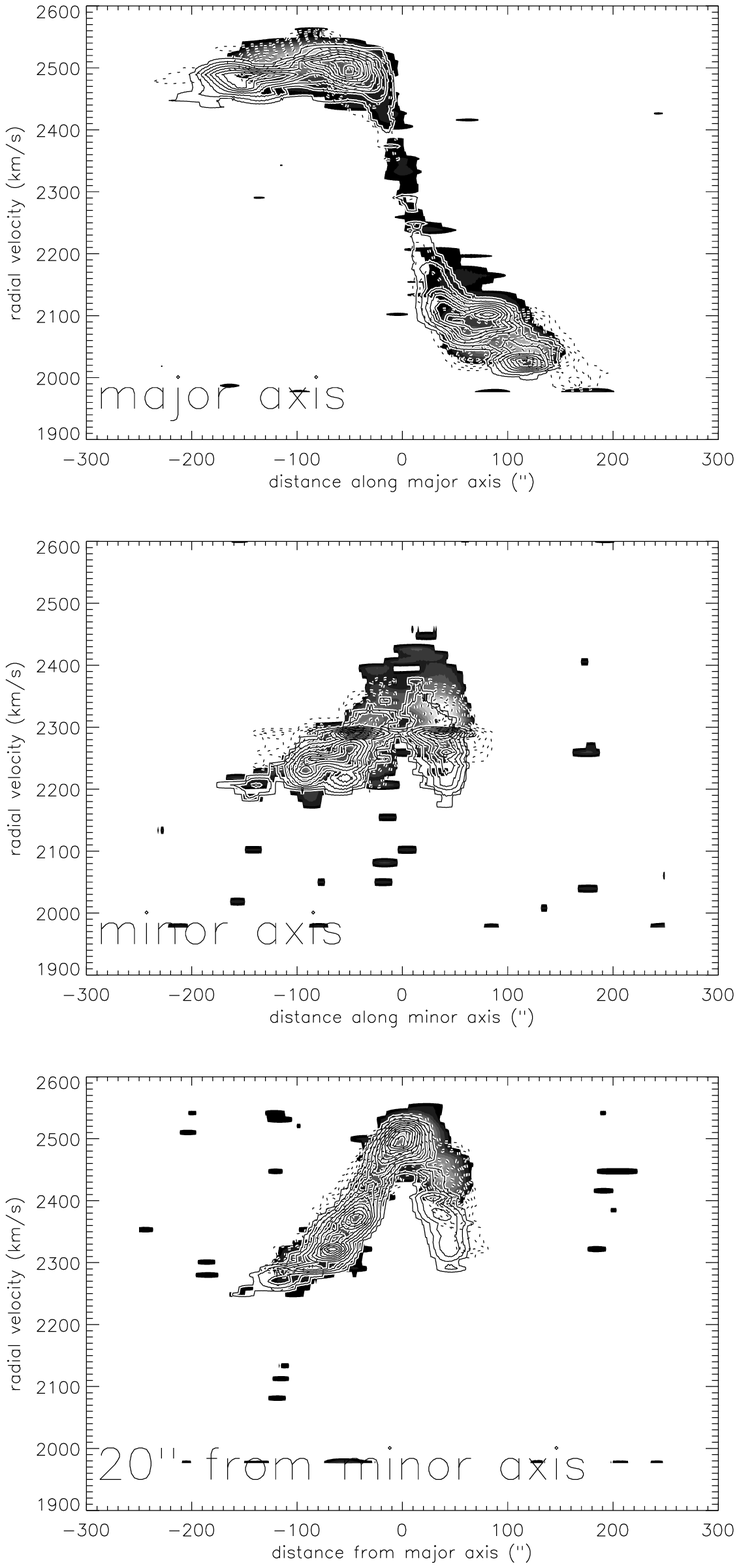}}
	\caption{Comparison between observed H{\sc i} (greyscale) and simulated position-velocity diagrams
	  for the LRP1 (dotted line) and LRP2 model (solid line).
	  Upper panel: cut along the major axis, northwest is to the right; 
	  middle panel: cut along the minor axis, northeast is to the left; 
	  lower panel: cut parallel to the minor axis $20''$ southeast from the galaxy center,
	  northeast is to the left.
	} \label{fig:n4501bb4test_cuts}
\end{figure} 


Both models also reproduce qualitatively the overall gas kinematics.
The two main characteristics of the observed H{\sc i} velocity field, the earlier flattening
of the rotation curve in the southeast and the regular velocity field in the
extended low surface-density northeastern region indicating rotation, are well reproduced.
This behavior is more pronounced in the LRP2 model snapshot. 
For a more detailed comparison we show position-velocity diagrams along the
major axis, minor axis, and parallel to the minor axis $20''$ southeast from the galaxy center
(Fig.~\ref{fig:n4501bb4test_cuts}).
Whereas the observed rising rotation curve to the northwest is reproduced,
the model rotation curve to the southeast decreases sightly, whereas the observed H{\sc i}
rotation curve stays constant.
The model nicely reproduces the H{\sc i} data for the cuts along and parallel
to the minor axis.
The LRP1 model distribution of the dispersion velocity reasonably agree with the
H{\sc i} observations. That of the LRP2 model shows an asymmetry along the major axis,
with a larger velocity dispersion to the southeast due to radial velocities
induced by ram pressure compression, which is not observed.

\subsection{The star formation distribution}

\begin{figure*} 
	\resizebox{\hsize}{!}{\includegraphics{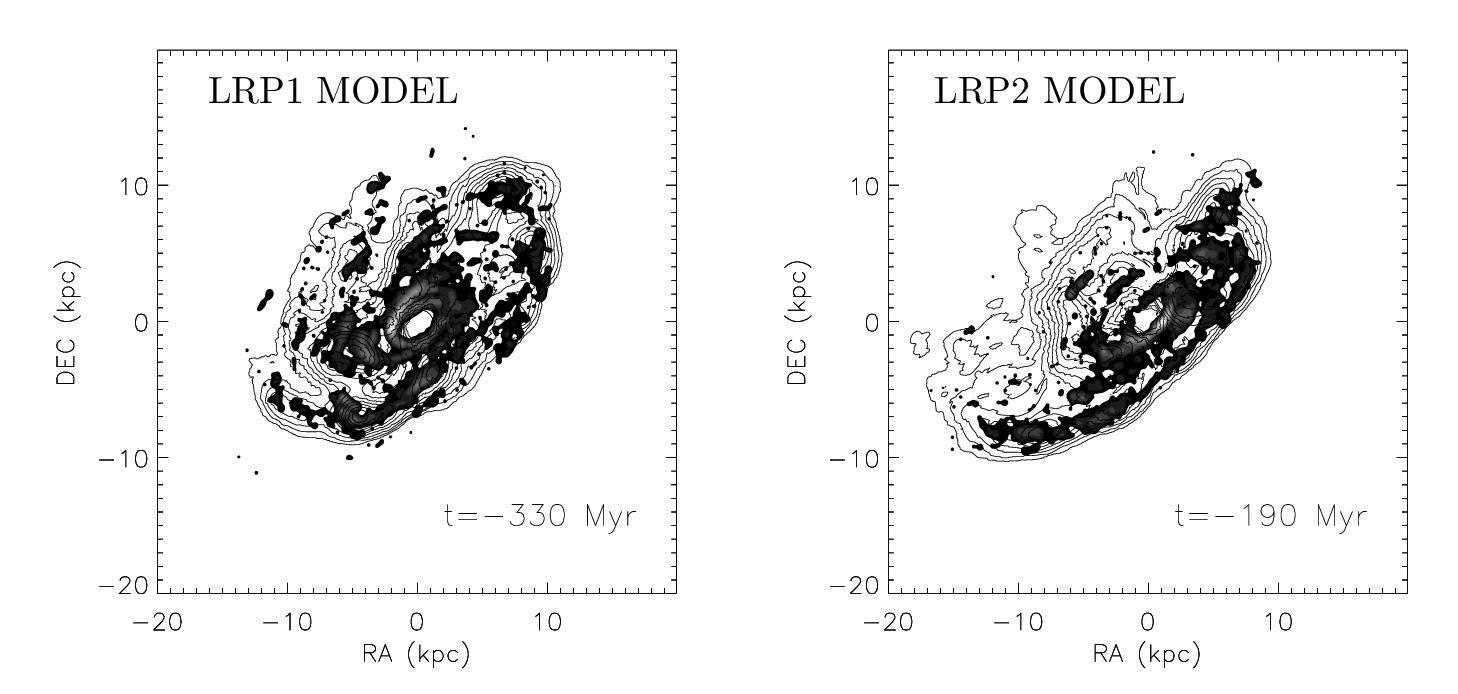}}
	\caption{Greyscale: model star formation distribution. Contours:
	  model gas distribution. The timestep is indicated in the lower right corner of each panel.
	} \label{fig:starform}
\end{figure*}

\begin{figure} 
	\resizebox{\hsize}{!}{\includegraphics{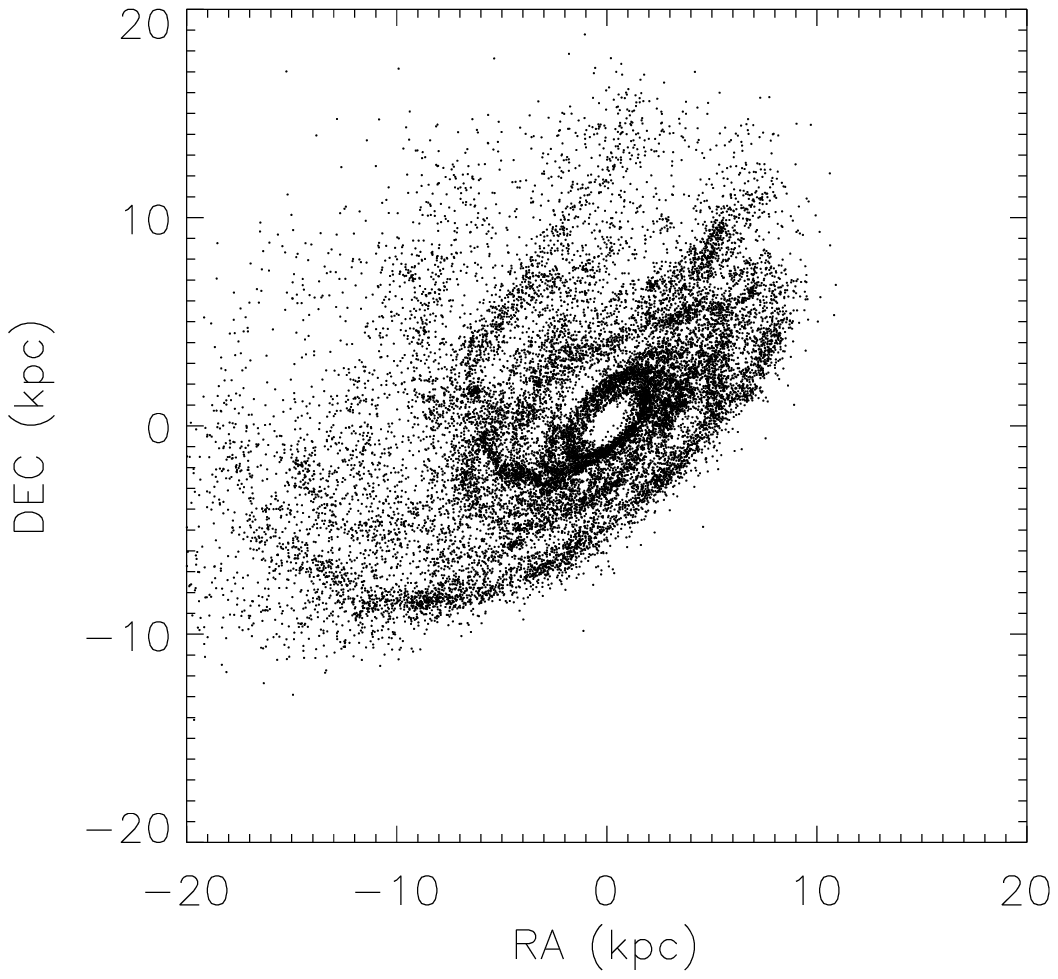}}
	\caption{Model cloud distribution at $t=-190$~Myr of the LRP2 snapshot.
	} \label{fig:lrpdist}
\end{figure}

The collisional component of the model consists of 20\,000 gas clouds of different mass
that can have inelastic collisions. For particles of equal size the collision rate,
i.e. the number of collisions per unit time, is ${\rm d}N/{\rm d}t = n \sigma v$ where
$n$ is the particle volume density, $\sigma$ the cross section, and $v$ the particle
velocity. The collisional timescale is $t_{\rm coll}=(n \sigma v)^{-1}$.
The mass involved in collisions per unit time and unit area is thus
$\dot{\Sigma}=\Sigma n \sigma v$, where $\Sigma$ is the surface density.
We now assume that the star formation $\dot{\Sigma_{*}}$ rate is proportional to the cloud 
collision rate $\dot{\Sigma_{*}} \propto \dot{\Sigma}$.
In a turbulent selfgravitating gas disk in hydrostatic equilibrium $n \propto \Omega^2$
and $\Sigma \propto \Omega$, where $\Omega$ is the angular velocity (Vollmer \& Beckert 2002,2003).
Thus, we obtain $\dot{\Sigma_{*}} \propto \Sigma^{1.5}$ which is close to the
observed Schmidt law (see, e.g. Wong \& Blitz 2002).
To compare the model star formation rates with H$\alpha$ observations we gather all collisions
up to $10$~Myr before the timestep of interest. The resulting model star formation rates
are shown in Fig.~\ref{fig:starform}.
These snapshots can be directly compared to Fig.~\ref{fig:HIHA}.
Both model snapshots show a ring of star formation in the inner 5~kpc
and enhanced star formation along the compressed ridge of high column density gas at the 
southwestern windward side. The northwestern H$\alpha$ spiral arm is less prominent in the model.
The observed lack of star formation at $\sim 1'$ east from the
galaxy center is only present in the LRP2 model. This is directly linked to the
discrepancy of the observed and simulated gas surface density at this location
discussed in Sec.~\ref{sec:compgas}. 
The underlying gas cloud distribution of the LRP2 model snapshot is shown in
Fig.~\ref{fig:lrpdist}. The observed faint outer H$\alpha$ arm in the west
of the galactic disk is not present in our model where the star formation peaks
together with the gas distribution. However, we observe a qualitatively similar behavior
of the LRP2 model star formation distribution southwest from the galaxy center:
there are two distinct arms with a clear gap between them. The external arm is
due to ram pressure compression, the internal arm is a spiral arm.  
Since the H$\alpha$ emission distribution also shows a double arm structure
in the southwest (Fig.~\ref{fig:HIHA}), 
we suggest that the southern part of the compressed H{\sc i} ridge
consists of two distinct narrow gas arms which are smeared out by the $30''$ resolution of the
H{\sc i} observations. We thus predict that future high resolution H{\sc i}
observations detect these two arms.

\subsection{The polarized radio-continuum emission}

\begin{figure*} 
	\resizebox{\hsize}{!}{\includegraphics{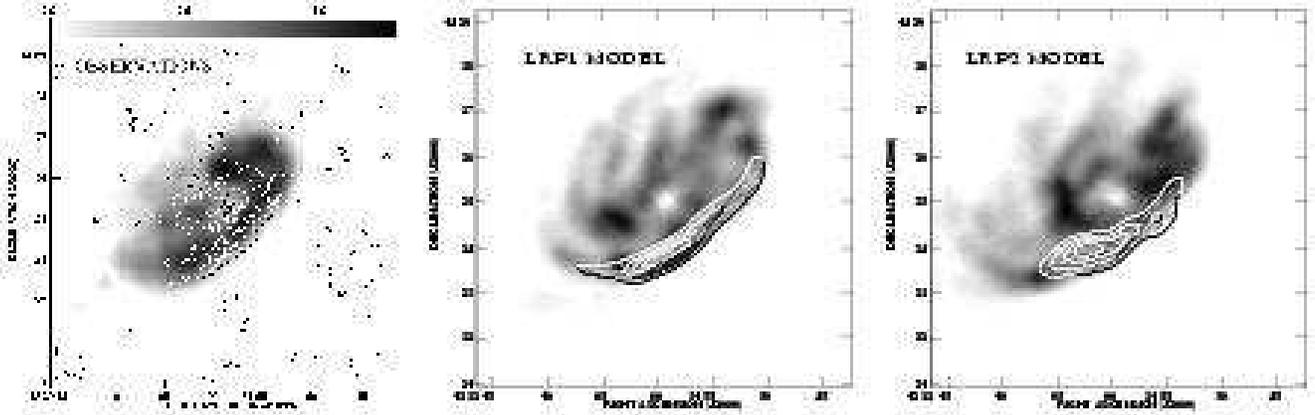}}
	\caption{Comparison between the observed and simulated gas surface density
	  distribution (greyscale; resolution: $30''$), the polarized radio-continuum emission
	  (contours; resolution: $18''$), and the sky-projected vectors of the large-scale
	  magnetic field.
	} \label{fig:HIPIcomparison}
\end{figure*}

The comparison between the observed and simulated gas surface density
distribution, the polarized radio-continuum emission, and the sky-projected vectors of the 
large-scale magnetic field is shown in Fig.~\ref{fig:HIPIcomparison}.
As explained in Sec.~\ref{sec:mhdmodel} we lose the large-scale magnetic field
in the inner disk due to numerical diffusion. 
Moreover, the observed polarized emission in the inner disk could be due to regular fields 
or compressed random fields due to dynamo action or compressed fields due to density waves, 
which are not part of the model. The model polarized emission is due to
an enhancement of the large-scale magnetic field by
shear and radial motions (induction equation) which are well present in the
model velocity fields. Therefore, while the loss of magnetic field energy by
diffusion is a numerical effect, the gain of magnetic field energy, and thus the
enhancement of polarized radio-continuum emission, is real.
We successfully reproduce the enhanced polarized radio-continuum distribution at the edge of 
the compression region. 
As observed the model polarized emission peaks in front of the H{\sc i} emission
for the LRP1 model. However, the south-eastern maximum of the LRP2 model is located
at the backside of the H{\sc i} maximum. The magnetic field vectors 
are parallel to this edge for the LRP1, in agreement with observations, whereas they
deviate from this direction at the location of the southeastern maximum of the LRP2 model.

\section{Discussion \label{sec:discussion}}

%
The first set of simulations with the temporal ram pressure profiles used in
Vollmer et al. (2001, 2005, 2006) qualitatively reproduce the observed H{\sc i} surface 
density distribution (LRP of Fig.~\ref{fig:n4501_p20} and HRP of Fig.~\ref{fig:n4501_p50}).
However, the deduced timescale to peak ram pressure of $120$~Myr is too short for the
galaxy to reach a position near enough to the cluster center where the expected ram pressure
corresponds to the model peak ram pressure. Therefore, we made a second set of
simulations where we doubled the widths of the temporal ram pressure profiles.
Since the inferred motion of the galaxy within the intracluster medium
does not permit it to reach a position where the peak ram pressure is strong
(as for NGC~4438 where $p_{\rm max}=5000$~cm$^{-3}$(km\,s$^{-1}$)$^{2}$, HRP1/2),
we prefer the simulations with  moderately strong peak ram pressure
($p_{\rm max}=2000$~cm$^{-3}$(km\,s$^{-1}$)$^{2}$, LRP1/2).
These dynamical simulations qualitatively reproduce the observed H{\sc i} surface density 
distribution and kinematics (Fig.~\ref{fig:momcomparison}), 
the observed massive star formation distribution
(traced by H$\alpha$ emission; Fig.~\ref{fig:starform}), and the 6~cm polarized radio
continuum emission distribution (Fig.~\ref{fig:HIPIcomparison}).

Within the LRP1/2 model the
current ram pressure on NGC~4501 is 400~cm$^{-3}$(km\,s$^{-1}$)$^{2}$ (LRP1 model) 
and 800~cm$^{-3}$(km\,s$^{-1}$)$^{2}$ (LRP2 model). 
The direction of the model galaxy's three dimensional velocity vector is $(0.48,-0.44,0.76)$
($x$-axis: RA, positive to the west; $y$-axis: DEC, positive to the north;
$z$-axis: line-of-sight, positive for the far side).
Based on this geometry derived from our simulations, NGC~4501's orbital velocity
is 1500~km\,s$^{-1}$ for both model snapshots LRP1 and LRP2. 
A simplified Gunn \& Gott criterion
(Gunn \& Gott 1972) yields
\begin{equation}
\Sigma_{\rm ISM} \frac{v_{\rm rot}^{2}}{R} = \rho_{\rm ICM} v_{\rm gal}^{2}\ ,
\end{equation}
where $R$ is the stripping radius, $\Sigma_{\rm ISM}$ the H{\sc i} column density, $v_{\rm rot}$ the
rotation velocity, and $v_{\rm gas}$ the orbital velocity of NGC~4501.
This estimate assumes a smooth and static intracluster medium.
Taking an H{\sc i} column density of $\Sigma_{\rm ISM}=10^{20}$~cm$^{-2}$
and a stripping radius of 12.5~kpc, we derive an intracluster medium density of
$\rho_{\rm ICM}=10^{-4}$~cm$^{-3}$.
This is consistent with the intracluster medium density at the projected location of NGC~4501
derived from X-ray observations under the assumption of hydrostatic equilibrium
$\rho_{\rm ICM}=7\,10^{-5}$~cm$^{-3}$ (Schindler et al. 1999).

The LRP1 model better reproduces (i) the outer edge of the H{\sc i} distribution compared
to the optical isophots (Fig.~\ref{fig:n4501_HI_stars}), (ii) the symmetric distribution of the
velocity dispersion (Fig.~\ref{fig:momcomparison}), (iii) the distribution of
the polarized radio-continuum emission (Fig.~\ref{fig:HIPIcomparison}), and
(iv) the galaxy has more time to approach the cluster center ($320$~Myr to peak ram pressure).
On the other hand, the LRP2 model better reproduces (i) the asymmetric H{\sc i}
distribution along the major axis with a larger extent to the southeast, (ii) the steepening of the
rotation curve to the northwest and its flattening to the southeast
(Fig.~\ref{fig:n4501bb4test_cuts}), and (iii) the gap between two arms of star formation
at the western edge of the galactic disk (Fig.~\ref{fig:starform}).
The timescale to ram pressure maximum of $190$~Myr is smaller than that of the LRP1 model,
but still sufficient to reach a position in the cluster where the model peak ram pressure
can be reached. Based on these arguments we regard the timescale of $320$~Myr to
peak ram pressure as an upper limit.

\begin{figure}  
	\resizebox{\hsize}{!}{\includegraphics{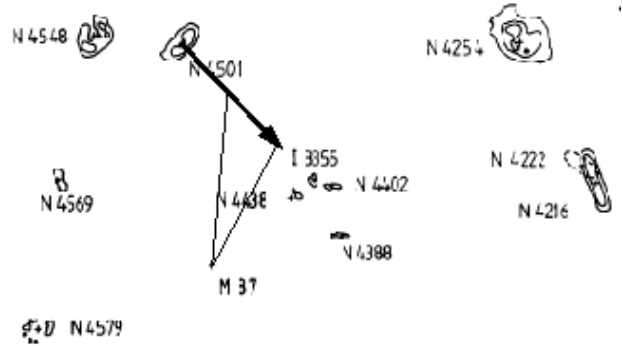}}
	\caption{Sketch of the linear extrapolation of 
	  NGC~4501's probable trajectory within the Virgo cluster.
	  Contours show the H{\sc i} surface density of the large spiral galaxies
	  around the cluster center (M87; from Cayatte et al. 1990).
	  The projected distance between NGC~4501 and M87 is $2^{\circ}$ (0.6~Mpc).
	  The two lines between M87 and the trajectory correspond
	  to the extrapolated galaxy's projected distance at peak ram pressure 
	  for the LRP2/HRP2 snapshots (the galaxy is observed 
	  $190$~Myr before peak ram pressure) and the LRP1/HRP1 snapshots
	  (the galaxy is observed $320$~Myr before peak ram pressure).
	} \label{fig:N4501inVirgo}
\end{figure}

Fig.~\ref{fig:N4501inVirgo} shows the linear extrapolation of NGC~4501's probable trajectory.
With a radial velocity with respect to the Virgo cluster mean of $1130$~km\,s$^{-1}$
a velocity within the sky plane of $\sim 1000$~km\,s$^{-1}$ is inferred. Thus, the galaxy
travels 0.33/0.19~Mpc from its present location within 320/190~Myr. At maximum ram pressure
the galaxy's trajectory is perpendicular to its distance vector to M87.
The projected angles between the distance vector to M87 and the galaxy's trajectory
are $112^{\circ}$/$125^{\circ}$ and the projected distances to M87 are
0.3~Mpc/0.48~Mpc, respectively. This translates into a three dimensional
distance to M87 of 0.39~Mpc and 0.54~Mpc at maximum ram pressure, respectively. The line-of-sight
distance from M87 at this moment is then 0.14~Mpc and 0.27~Mpc, respectively.
The present line-of-sight distance of NGC~4501 is 0.11~Mpc in front of M87 for the
LRP1 model ($\Delta t = -320$~Myr) and 0.13~Mpc behind M87 for the LRP2 model
($\Delta t = -190$~Myr).

This overall picture (see also table~\ref{tab:summary}) makes us conclude that 
the outer western ridge of 
star formation is induced by nearly edge-on ram pressure stripping ($i \sim 10^{\circ}$).
A comparable case is NGC~4654 (Vollmer et al. 2003), where nearly edge-on
ram pressure stripping ($i \sim 0^{\circ}$) leads to an enhanced
northwestern H$\alpha$ emission in the compression region.
This hypothesis can be verified by the
determination of the star formation history from multiband photometry including UV emission 
and optical spectra. If our scenario is correct we expect a recent small starburst
that created the western outer H$\alpha$ arm.

\section{Conclusions \label{sec:conclusions}}

VIVA H{\sc i} observations of the Virgo spiral galaxy NGC~4501 are presented.
The outer edge of the H{\sc i} distribution follows one optical isophot contour
except in the northeast, where a region of low surface-density gas is discovered.
Whereas the B band image is asymmetric, the lowest B band isophot contour and the
H band isophot contours are symmetric. The observed southwestern ridge of 6~cm
polarized radio-continuum emission (Vollmer et al. 2007) is located at the outer edge 
of the H{\sc i} distribution.
Both show the same rapid increase at the edge, which is expected in a ram pressure
scenario. Along the gas ridge, a faint H$\alpha$ emission ridge is present.
The maximum of the polarized radio-continuum emission are located within the
gap between the H$\alpha$ emission of the spiral arms and the faint emission ridge.

We compare these observational characteristics with detailed dynamical models
including MHD calculations. Models with different galaxy orbits
(high and low ram pressure) and disk-wind angles between the disk and the
orbital plane are presented. Model snapshots at different times from peak ram pressure
are compared to the multi-frequency observations. For the first time we
present model maps of massive star formation.
We qualitatively reproduce the gas distribution,
kinematics, star formation distribution, and polarized radio-continuum emission
distribution. We conclude that ram pressure stripping can account for the main characteristic.
The western ridge of enhanced gas surface density and enhanced
polarized radio-continuum emission is due to ram pressure compression. 
Assuming radial orbits in a static intracluster medium of the Virgo cluster
(Vollmer et al. 2001), NGC~4501 will have its nearest approach to the cluster core,
i.e. peak ram pressure, in $\sim 200-300$~Myr (Fig.~\ref{fig:dessin1}).
It is argued that the faint western H$\alpha$ emission ridge
is induced by nearly edge-on ram pressure stripping.

\begin{acknowledgements}
BV would like to thank the MPIfR (P.~Reich) for computational support.
This research has made use of the GOLD Mine Database.
This work was supported by Polish-French (ASTRO-LEA-PF)cooperation program,
and by Polish Ministry of Sciences grant PB 378/P03/28/2005 and PB 2693/H03/2006/31.
We would like to thank the referee for his stimulating comments.
\end{acknowledgements}

\end{document}